# Superconductivity and Ferroelectric Orbital Magnetism in Semimetallic Rhombohedral Hexalayer Graphene

*Jinghao Deng[1#], Jiabin Xie[1#], Hongyuan Li[1#], Takashi Taniguchi[2], Kenji Watanabe[3], Jie Shan[4,5], Kin Fai Mak[1,5], Xiaomeng Liu[1*]*

*[1]Laboratory of Atomic and Solid State Physics, Cornell University, Ithaca, NY, USA*
*[2]Research Center for Materials Nanoarchitectonics, National Institute for Materials Science, 1-1 Namiki, Tsukuba 305-0044, Japan*
*[3]Research Center for Electronic and Optical Materials, National Institute for Materials Science, 1-1 Namiki, Tsukuba 305-0044, Japan*
*[4]School of Applied and Engineering Physics, Cornell University, Ithaca, NY, USA*
*[5]Max Planck Institute for the Structure and Dynamics of Matter, Hamburg, Germany*

*[#]These authors contributed equally*
**Correspondence author E-mail: xl956@cornell.edu*

**Abstract: Rhombohedral multilayer graphene has emerged as a promising platform for exploring correlated and topological quantum phases, enabled by its Berry-curvature-bearing flat bands. While prior work has focused on separated conduction and valence bands, we probe the extensive semimetallic regime of rhombohedral hexalayer graphene. We survey a rich phase diagram dominated by flavor-symmetry breaking and reveal an electric-field-driven band inversion by fermiology. Near this inversion, we observe two superconducting-like states confined to regions with coexisting electron and hole Fermi surfaces, suggesting a possible dual-carrier origin. In addition, we identify a ferroelectric orbital magnet that undergoes sharp switching under unipolar electric fields, signaling spontaneous electric polarization. Unlike previously reported multiferroicity near zero electric field, this new state exhibits magnetic hysteresis reversible by electric field, consistent with a phenomenological model of coupled electric and magnetic polarization. Our work elucidates the correlated semimetal regime in rhombohedral graphene and underscores its potential to host diverse quantum phases.**

Rhombohedral multilayer graphene has emerged as a versatile platform for exploring strongly correlated and topological quantum phases, owing to its unique flat bands with nontrivial Berry curvature[1,2]. In the rhombohedral stacking configuration, each graphene layer is laterally displaced by one carbon-carbon bond relative to the one beneath (Fig. 1a, inset). This stacking pattern leads to strong interlayer hybridization that gaps out electronic states on the inner layers (black dots), while leaving only the top and bottom sublattices (blue and red dots) electronically active[3,4]. As the number of layers $N$ increases, the growing spatial separation between these active sites suppresses their hybridization, giving rise to nearly dispersionless (flat) bands at band edges. These atomic-scale flat bands greatly enhance the role of electron–electron interactions, making the system highly susceptible to a wide range of many-body instabilities. Indeed, correlated phenomena such as spontaneous symmetry breaking of spin and valley[5–10], superconductivity[6,11–18], and nematic order[12,19] have been reported in this system.

Beyond strong correlations, the flat bands in rhombohedral graphene also carry a nonzero Berry curvature, originating from the winding of pseudospin textures. In each valley, the total Berry phase scales as $\pi N$[1,3,4]. While time-reversal symmetry ensures cancellation of Berry phases between valleys, spontaneous valley polarization can lift this cancellation and produce orbital ferromagnetism. This mechanism underlies the recent observations of multiferroic orbital ferromagnetism[10], the quantum anomalous Hall effect[20,21], the long-sought fractional quantum anomalous Hall state[22,23], and potential chiral superconductivity[24].

To date, most studies of rhombohedral graphene have focused on single-band phenomena under large perpendicular electric fields that open a band gap. Here, we investigate the semimetallic regime of hexalayer rhombohedral graphene, where conduction and valence bands overlap. Compared with thinner structures ($N \leq 5$), the six-layer system exhibits pronounced band overlap that creates an extended semimetallic region, while its flatter low-electric-field band dispersion amplifies correlation effects (Fig. 1c). Such two-band systems are of fundamental interest for both topology and correlations: topological insulators can arise from band inversion between orbitals[25,26], and interband electron–hole pairing can give rise to excitonic insulators[27]. The tunable band overlaps of rhombohedral graphene thus provide fertile ground for correlated semimetal physics.

We report a rich phase diagram in this regime, featuring correlated insulators, superconductivities, two distinct multiferroic orbital magnetic orders, amid an intricate network of phase boundaries (Fig. 1e). Quantum oscillation fermiology reveals sweeping two-fold symmetry breaking, links the phase boundaries to Lifshitz transitions of the two-flavor band structure, and uncovers a band inversion in one flavor as the electric field is tuned. Around the inversion, we identify two superconducting states, SC1 and SC2, each occurring in a distinct dual-carrier regime and exhibiting in-plane critical fields far exceeding the Pauli limit. SC1 emerges from two emergent Fermi pockets that display pronounced temperature sensitivity, pointing to either a temperature-driven Fermi-surface reconstruction or a substantially enhanced effective mass.  In addition, within the semimetallic regime, we uncover an orbital magnetic state exhibiting both electric- and magnetic-field hysteresis. Strikingly, the polarity of the magnetic hysteresis is reversed across a critical electric field value, signaling an intertwined electric–magnetic order. Our understanding of symmetry breaking and Lifshitz transitions in this system, established through detailed quantum oscillation fermiology, can guide future exploration of higher $N$ rhombohedral graphene.

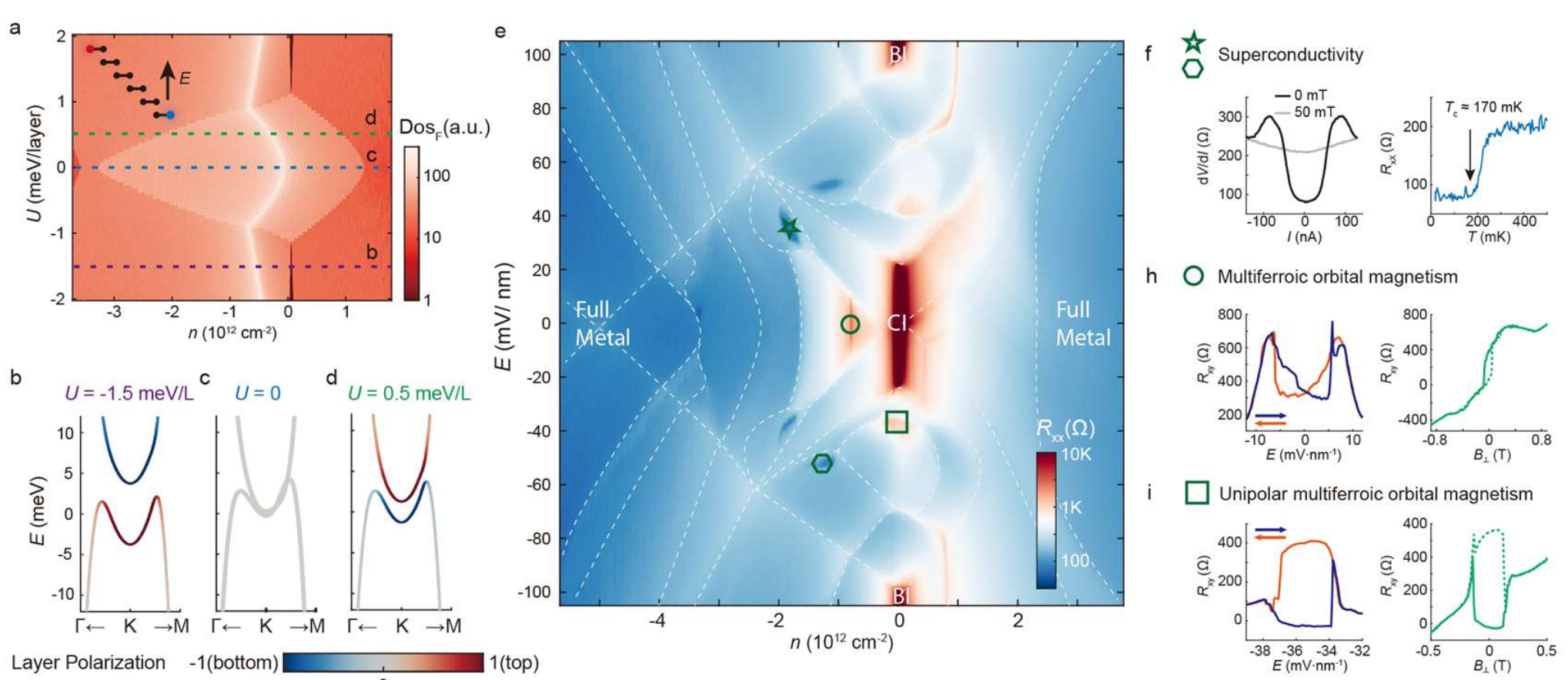

***Figure 1 | Band structure and phenomenology of semimetallic rhombohedral hexalayer graphene.***
***a,*** *Density of states at the Fermi energy vs carrier density n and interlayer potential U, from a tight-binding calculation. The spindle-shaped pink region indicates simultaneous intersection of the Fermi energy with conduction and valence bands. Inset, schematic of the rhombohedral six-layer structure, with active sublattices highlighted in blue and red.* ***b,*** *Band structure near the K point for three U values marked in* ***a****, with line color denoting layer polarization.* ***e,*** *Longitudinal resistance vs n and perpendicular electric field E (pointing upward). Major features are outlined with dashed lines; BI and CI denote band and correlated insulators. Symbols mark locations of selected quantum phases.* ***f,*** *Superconductivity signatures of SC1 (the green star symbol), including a magnetic-field–sensitive differential-resistance gap and a sharp resistivity drop near 170 mK. Similar superconducting behavior is observed* *at the green hexagon symbol (SC2, see Fig. S11).* ***h****, Ferrovalley orbital magnetism in the triangular region around E=0, switchable by electric and magnetic fields.* ***i****, Ferroelectric orbital ferromagnetism with a unipolar electric-field hysteresis loop.*

We first review the single particle bands. Figure 1a shows the calculated density of states at the Fermi energy ($DOS_F$) as a function of carrier density and interlayer potential $U$, obtained from a tight-binding model. $U$ represents the onsite potential difference between adjacent layers and is equivalent to the experimentally applied perpendicular electric field $E$. At $U = 0$ (Fig. 1c), the electron and hole bands overlap significantly, with the corresponding wavefunction equally distributed between the top and bottom layers. Applying a finite electric field breaks this symmetry, leading to layer-polarized states at the band edges, with the polarization direction depending on the field orientation (Figs. 1b & 1d). As the field increases, the potential further separates the conduction and valence bands and produces a band gap (Fig. 1d). Across electric fields, the conduction band is approximately a parabola while the valence band is shaped as an inverted Mexican hat.

Figure 1e shows experimental four-terminal longitudinal resistance (on a logarithmic scale) as a function of carrier density $n$ and perpendicular electric field $E$ (pointing upwards). Consistent with the calculated density of states at the Fermi level in Fig. 1a, a transport gap opens at charge neutrality once the electric field exceeds a threshold of approximately 100 mV/nm. Below this threshold, a rich landscape of phase boundaries (delineated by dashed lines) and quantum phases unfolds in the single particle semimetal regime.

At charge neutrality near zero electric field, situate the correlated insulator (CI, temperature dependence shown in Fig. S1). Slight hole doping brings a triangular region between $n = 0$ and $n = -1.8\times10^{12}$ $cm^{-2}$ that is symmetric about zero $E$ that hosts orbital ferromagnetism, characterized by hysteretic Hall response in magnetic and electric fields (Fig. 1h, more in Fig. S2), similar to those seen in pentalayer rhombohedral graphene[10]. A region at the tail of this triangle holds a superconducting-like phase (green star in Fig. 1e, which we refer as SC1). Although the resistance does not reach zero, its temperature, magnetic field, and current dependence are consistent with superconductivity or a fragile superconductor (Fig. 1f). Around $n = 0$ and $E \approx -35$ mV/nm, we identify a second multiferroic orbital ferromagnetic phase that exhibits unipolar switching behavior in the electric field (Fig. 1i). That is, both forward and backward switching is offset in the same electric field direction (~ –34 mV/nm and ~ –37 mV/nm, respectively). Near the symmetry-breaking phase boundary extended from this second orbital magnetism state, we identify an additional superconducting-like phase (green hexagon in Fig. 1e, which we refer as SC2).

Understanding the correlated insulator is central to interpreting the phase diagram. At $E = 0$, single-particle theory predicts wavefunctions as equal-weight superpositions of the top and bottom layers (Fig. 1c). Electron interactions, however, drive pseudospin ferromagnetism that produces spontaneous layer polarization and opens an emergent gap[28]. This polarization can orient upward or downward, effectively mimicking a positive or negative external electric field, which we refer to as a positive or negative emergent gap.

In rhombohedral graphene, a uniform emergent gap across spin and valley would place all four bands in the same layer, incurring a large Coulomb penalty. At $E = 0$, the energetically favored state instead adopts a two-by-two splitting: two isospins acquire positive emergent gaps and the other two acquire negative ones, thereby forming two flavored bands (Fig. 2j). The specific flavor pairing determines the broken symmetry—spin, valley, or a mixed combination. The commonly recognized layer antiferromagnetism corresponds to spin splitting, where spin-up and spin-down experience opposite emergent fields and occupy opposite layers[2,8,18,29,30]. Alternatively, an inter-valley coherent insulator can arise from superpositions of the $K$ and $K'$ valleys with the canceling Berry phase[31,32]. Without the ability to distinguish them, we refer to the two flavored bands, each two-fold degenerate, generically as flavor A (negative emergent gap) and B (positive emergent gap).

With increasing electric field, the gap at flavor A shrinks while that of flavor B expands, reflecting the lowering (raising) of potential energy in the bottom (top) layer (Figs. 2i&j). At larger fields, the gap at flavor A closes and inverts from negative to positive (Figs. 2g&h). As we show below, the phase diagram of rhombohedral hexalayer graphene (Fig. 1e) is dominated by two-fold degenerate half metals that can be understood in connection with this flavor-broken band inversion.

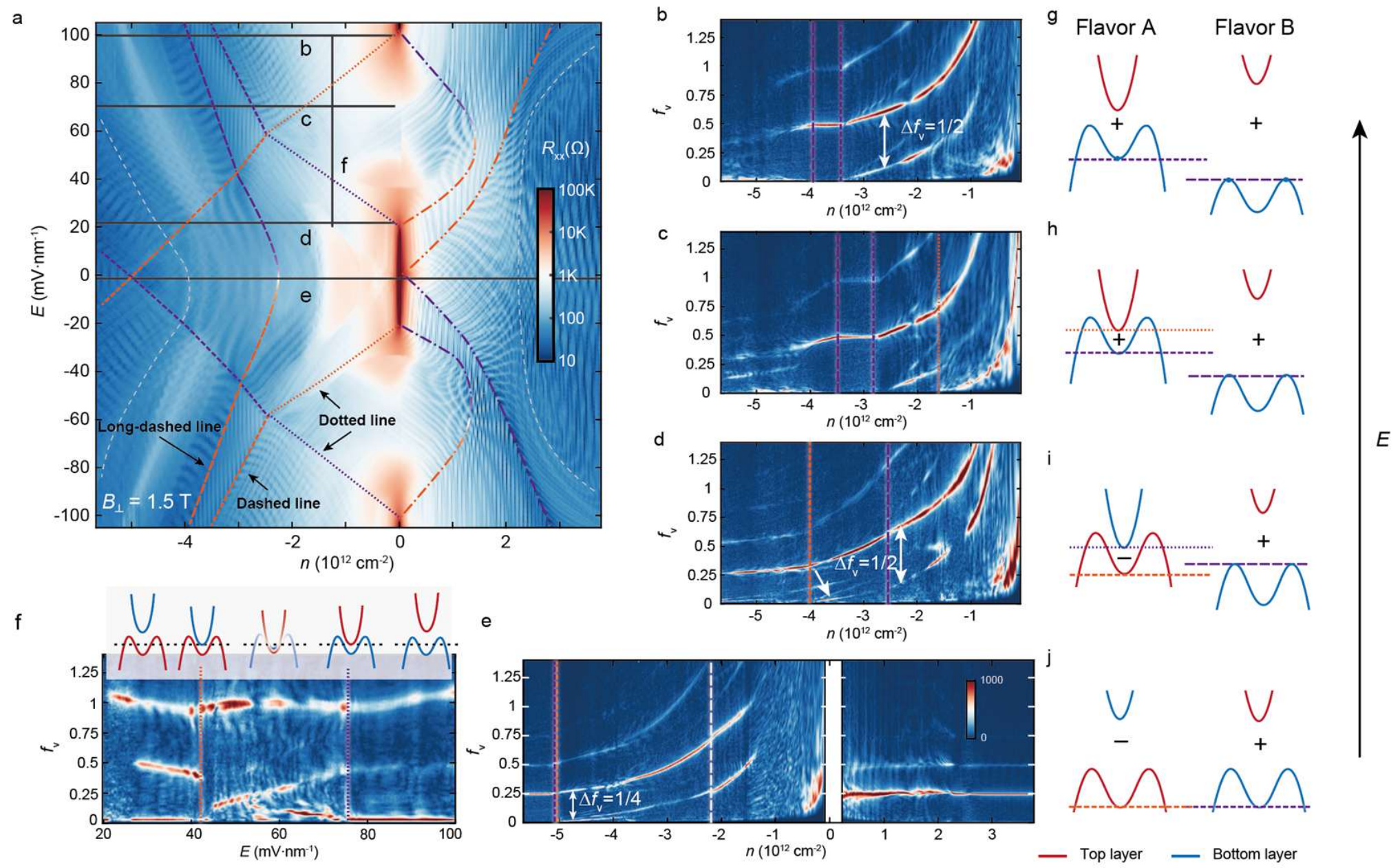

***Figure 2 | Fermiology of symmetry-breaking semimetal.***
***a,*** *Longitudinal resistance vs n and E at 1.5 T. Purple and orange lines mark phase boundaries associated with the discussed Lifshitz transitions, with colors reflecting the inferred layer polarization of the corresponding band.* ***b–e,*** *Fourier transform of $R_{xx}(1/B)$ vs n and normalized frequency along the electric-field line cuts indicated in* ***a****. Vertical purple and orange lines denote quantum-oscillation changes corresponding to a Lifshitz transition, aligned with the phase boundaries in* ***a****.* ***f,*** *Quantum-oscillation map vs E along the vertical cut in* ***a****, with the band diagram above showing the corresponding evolution of flavor A. Black dotted lines represent the Fermi energy.* ***g–j****, Schematic band diagrams at various electric fields matching* ***b–e****. Dotted and dashed lines indicate Fermi levels at the Lifshitz transitions, with line patterns consistent with previous panels.*

In Fig. 2a, we show the longitudinal resistance under a perpendicular magnetic field of 1.5 T. Sudden changes in the quantum oscillations occur along the orange and purple boundaries, which we identify as Lifshitz transitions. These phase boundaries form a subset of those in Fig. 1e. Here we focus on the hole-doped side, while the electron-doped fermiology is presented in Fig. S3. Figs. 2b–f show the fast Fourier transform (FFT) of quantum oscillation along horizontal and vertical line cuts of Fig. 2a (raw data are shown in Figs. S4). The vertical axis in these panels, $f_v$, is defined as the normalized oscillation frequency: $f_v = f_{FFT} \cdot e/nh$, where $f_{FFT}$ is the oscillation frequency of $R_{xx}$ ($1/B_\perp$), and $e$, $n$, and $h$ are the electron charge, carrier density, and Planck's constant, respectively. Physically, $f_v$ represents the number of Bloch states enclosed by the Fermi surface relative to the carrier density. For a simple, non-degenerate parabolic band, $f_v$ peaks at 1, while double degenerate parabolic bands produce a peak at 0.5.

In Fig. 2b, to the right of the purple dashed line, two dominant oscillation frequencies appear, separated by 0.5 in $f_v$, indicating an annular Fermi surface with two-fold degeneracy. This observation illustrates that doping the band insulator creates symmetry-breaking metal, consistent with flavored bands in Fig. 2g. The lower-frequency branch vanishes at the dashed purple line, signaling the disappearance of the inner circle of the annular Fermi surface. Below this line, the Fermi surface becomes simply connected, giving rise to a peak fixed at $f_v = 0.5$. As hole doping increases further, the Fermi energy drops to the band edge of flavor B—marked by the long dashed purple line in Figs. 2b and 2g—entering a regime of partial flavor polarization (where the sum rule of $f_v$ is violated, discussion of the sum rule is presented in Fig. S5).

At reduced electric field, Fig. 2c shows the same phase boundaries (purple dashed lines), shifted in the positive $n$ direction. This reflects an upward energy shift in the corresponding band features with decreasing electric field, consistent with bottom-layer polarization. This trend is also evident in the backward tilt of the purple lines in the *n*-*E* map of Fig. 2a. Additionally, a new electron pocket emerges in Fig. 2c and vanishes near $n \sim -1.5\times10^{12}$ cm$^{-2}$, bounded by the dotted orange line. To the right of the dotted orange line lies a flavor-polarized semimetal with an annular hole Fermi surface and a circular electron Fermi surface. Its existence corroborates the band overlap between conduction and valence bands illustrated in Fig. 2h. The dotted orange line marks the conduction band edge; its correspondence in *n*-*E* map (Fig. 2a) has a positive slope that indicates top-layer polarization. Accordingly, we associate purple lines with bottom-layer band features and orange lines with top-layer polarized bands.

As the electric field lowers further (< 60 mV/nm), we identify similar phase boundaries, marked by orange dashed lines and purple dotted lines, corresponding to the valence band minimum and conduction band edge (white arrow in Fig. 2d and $f_3$ in Fig. S5i). Yet, the energy evolution of these boundaries in the electric field is opposite to that of $E$ > 60 mV/nm, evident from their line tilts in Fig. 2a. This indicates an inversion between conduction band edge (dotted line) and valence band bottom (dashed line) as shown in flavor A in Figs. 2h&i.

An electric-field line cut at $n = -1.26 \times 10^{12}$ cm$^{-2}$ captures the inversion process (Fig. 2f). To the right of the dotted purple line, we observe annular Fermi surfaces polarized to the bottom layer. As $E$ decreases past the purple boundary (~ 75 mV/nm), a top-layer–polarized electron Fermi surface emerges. With further reduction of $E$, this top-layer surface expands while the bottom-layer surface shrinks, consistent with their relative potential shifts (Fig. 2f, upper inset). At $E$ ~ 42 mV/nm, the two Fermi surfaces cross; below this point, the bottom-layer polarized band becomes the electron band and rises above the Fermi level at the dotted orange line. The band inversion is then complete, leaving a top-layer–polarized valence band.

Most boundaries in Fig. 1e can be understood by such an $E$-dependent but $n$-independent band picture. However, notable exceptions arise. At large electron and hole doping (grey dashed lines in Fig. 2a), the bands regain fourfold degeneracy (Figs. 2d&e). More puzzling is the observation at $E$ = 0: to the right of the long-dashed white line in Fig. 2e, the annular Fermi surfaces are only two-fold degenerate, in contrast to the band picture of Fig. 2j, which predicts fourfold degenerate bands at $E$ = 0. Equally striking, the valley-polarized orbital magnetism at zero electric field, marked by the green circle in Fig. 1e, is incompatible with the widely assumed layer antiferromagnet at charge neutrality under the same electric field[8,30]. Together, these anomalies suggest either evolution of the correlated band upon doping or a ground state with an alternative flavor polarization.

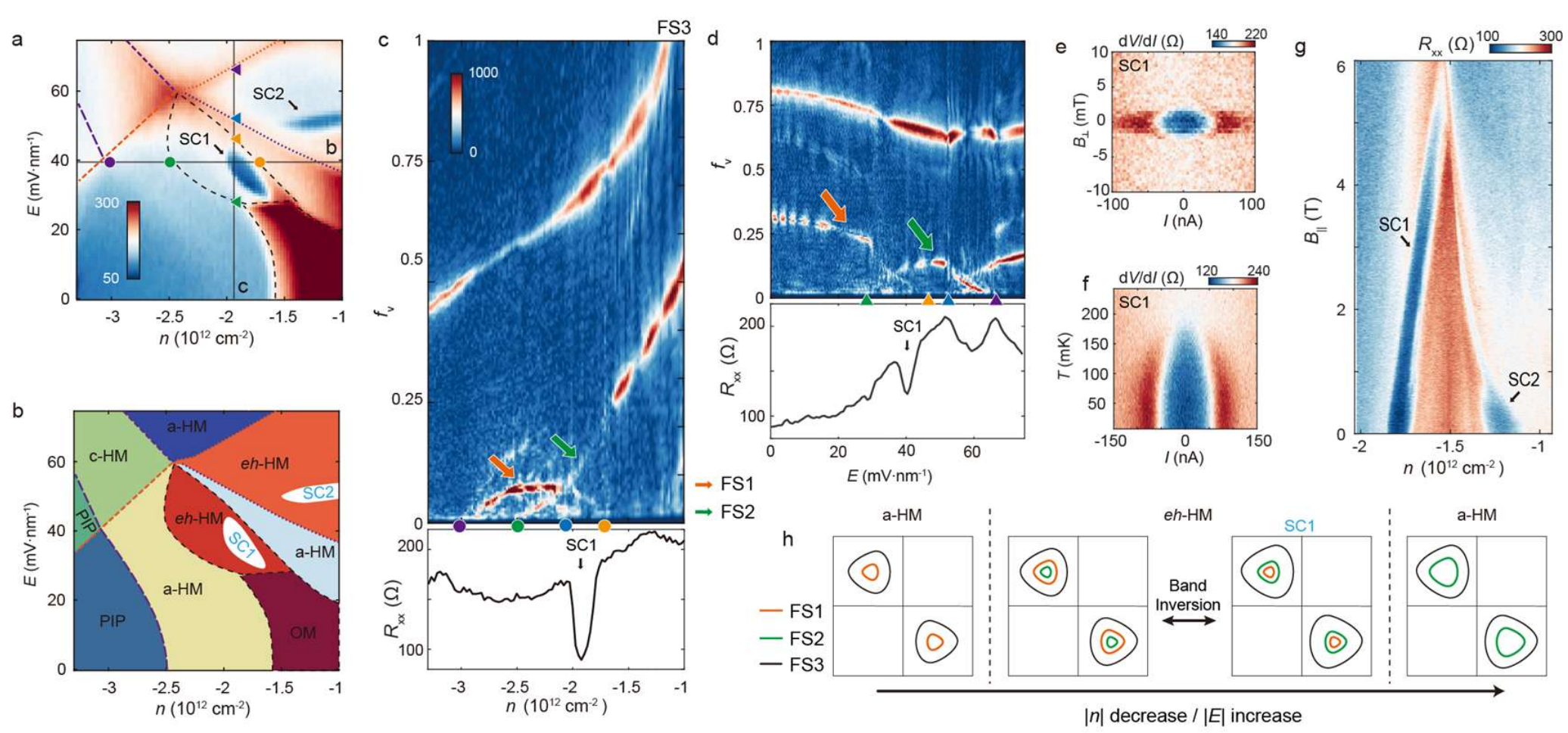

***Figure 3 | Emergent Fermi surface and superconductivity.***
***a,*** *Longitudinal resistance in a narrow n-E region near band inversion.* ***b,*** *Schematic phase diagram derived from experimental quantum oscillation results with labels. a-HM, half metal with an annular hole-like Fermi surface; c-HM, half metal with a circular hole-like Fermi surface; eh-HM, dual-carrier half metal with an annular hole-like* *and an additional electron-like Fermi surface; OM, orbital magnetism states; PIP, partially isospin polarized states; SC1 and SC2, superconductivity states.* ***c&d,*** *Fermiology (top panel) and zero magnetic field longitudinal resistance (bottom panel) along the horizontal and vertical grey lines in* ***a****; dots and triangles mark $f_\nu$ peaks reaching zero frequency, with corresponding n–E positions indicated in* ***a****. Orange and green arrow marks FS1 and FS2, respectively.* ***e,*** *Differential resistance of SC1 vs current and magnetic field at base temperature.* ***f,*** *Differential resistance of SC1 vs current and temperature at zero field.* ***g,*** *In-plane field dependence of longitudinal resistance along a line cut through two superconducting phases (see Fig. S11a); the main phase corresponds to the left blue strip.* ***i,*** *Schematic illustration of Fermi surface evolution near the eh-HM region. Each pair of Fermi contours occupied two blocks of a four-panel grid, showing their two-fold degeneracy nature among four different possible spin-valley states.*

Most interestingly, we observe an intriguing Fermi surface reconstruction around the intersection of the purple and orange dashed and dotted lines, where the bands in flavor A invert. Figures 3a and 3b highlight the intricate features in this region. Between the multiferroic triangle (bottom right) and the band crossing intersection, faint features in $R_{xx}$, outlined by black dashed lines, encircle an area that resembles a leaf.

Figures 3c and 3d present quantum oscillation fermiology along horizontal and vertical line cuts across the leaf–shaped region (cut positions indicated by solid grey lines in Fig. 3a). Along the horizontal cut (Fig. 3c), a Fermi surface—denoted FS1—emerges as $n$ reaches the purple dot, grows with increasing $n$, and then shrinks and disappears near the orange dot, forming an arc. Such evolution is incompatible with a density-independent band structure, as it would require the Fermi level to enter and then completely fill a small, isolated band that is detached from the surrounding valence-band continuum, which we consider unlikely in the present system. Inside this arc, a separate electron-like Fermi surface emerges near the green dot, which we call FS2. Notably, both FS1 and FS2 exhibit strong temperature sensitivity, becoming completely suppressed at 500mK (Fig. S6). This suggests either they are low-temperature emergence features or a substantially enhanced effective mass (Fig. S7). In addition, a higher-frequency peak, FS3, exceeds the sum of FS1 and FS2 by 0.5 (Fig. S5), indicating twofold degeneracy. Between the green and orange dots, all three Fermi surfaces coexist, corresponding to annular hole pockets and a circular electron pocket (Fig. 3h).

The fixed-$n$ line cut in Fig. 3d reveals a complementary evolution with the electric field. An arc-shaped feature spans the green and purple triangles, with additional Fermi-surface terminations at the yellow and green triangles, consistent with two successive band inversions. The inversion near 60 mV/nm is analogous to that shown in Fig. 2f, while the inversion near 40 mV/nm corresponds to a band inversion between FS1 and FS2, similar in nature to the carrier-density-driven crossing observed in Fig. 3c (see Fig. 3h). This line cut therefore captures two distinct dual-carrier Fermi-surface regimes: one between the green and yellow triangles, and another between the blue and purple triangles.

To clarify the relationship between resistance features and fermiology, we mark Lifshitz transition points identified from Figs. 3c and 3d onto Fig. 3a. These symbols closely align with marked phase boundaries. A comprehensive set of fermiology measurements tracking Fermi-surface evolution as a function of electric field is presented in Figs. S8–S10, with a summary shown in Fig. 3b. Notably, we identified two electron-hole half-metal states (*eh*-HM), one corresponding to the leaf-shaped region, and the other to the broader semimetallic regime spanning the correlated and band insulators.

The two superconducting states, SC1 and SC2, are embedded within these electron-hole half-metal regions (Fig. 3b-d). In addition to a sharp drop in resistance (Figs. 1f and S11), they exhibit characteristic features of superconductivity in differential conductance and under out-of-plane magnetic field (Figs. 1f, 3e, 3f, and S11). SC1 persists up to an in-plane magnetic field of approximately 6 T—more than an order of magnitude higher than the estimated Pauli limit of ~300 mT based on the observed $T_c$ (see Fig. S12 for resistance $n$-$E$ maps at various in-plane magnetic fields). The resistance does not fully reach zero in either superconducting-like state, which may be due to inadequate filtering or may indicate a fragile superconducting state.

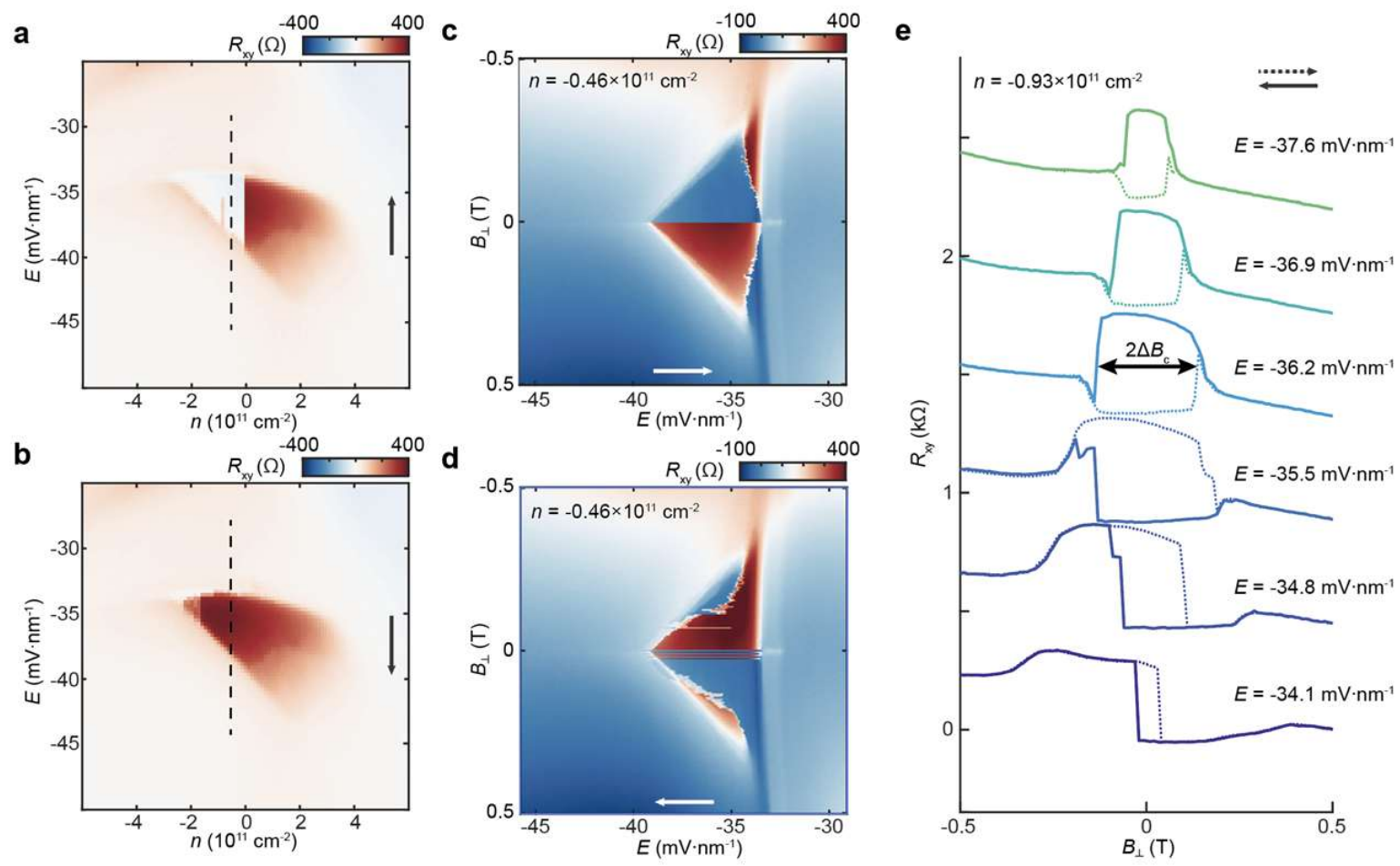

***Figure 4 | Ferroelectric orbital magnetism.***
***a&b,** Hall resistance in a small n-E area around the green square mark of Fig. 1e under a small magnetic field of $B_\perp$ = 50 mT. The fast scan direction is indicated by the arrow inside each panel. **c&d**, Electric field hysteresis vs perpendicular magnetic fields. The direction of scanning is indicated by white arrows. **e,** Magnetic field hysteresis vs E, with dashed (solid) lines plotting forward (backward) scan.*

We next discuss a novel multiferroic order that emerges in a comet-shaped region of the $n$-$E$ phase space (Fig. 1e, Figs. 4a&b). This region is located near $n = 0$ at finite $E$ within the semimetal regime. On the left of this region, a sharp hysteresis in Hall resistance $R_{xy}$ is observed when sweeping the electric field under a 50 mT stabilizing perpendicular magnetic field (Fig. 1e, Figs. 4a&b). The step-like $R_{xy}$ vs $E$ (Fig. 1i) differs qualitatively from the butterfly-shaped $E$-hysteresis seen in the triangle region near $E = 0$ (Fig. 1h) and previously reported in pentalayer graphene[10]. The sharp switching is indicative of spontaneous electric polarization. We therefore refer to this state as ferroelectric orbital magnetism (see Figs. S13 and S14 for its carrier density and temperature dependency). The polarity of this electric-field hysteresis reverses as $B_\perp$ changes sign, while the size of the hysteresis loop decreases with increasing $B_\perp$ magnitude (Figs. 4c&d).

The magnetic-field hysteresis of $R_{xy}$ exhibits even more striking behavior. At a fixed $n = -0.93 \times 10^{11}$ cm$^{-2}$, the polarity of the magnetic hysteresis reverses as the electric field is tuned: for $E \geq -35.5$ mV/nm, forward $B$-field sweeps (dashed lines) yield higher $R_{xy}$ than backward sweeps (solid lines), whereas for $E \leq -36.2$ mV/nm the opposite is true. This demonstrates that the orbital momentum does not always align with external field. Such behavior contrasts sharply with the multiferroic orbital magnetism previously reported in pentalayer graphene, where the magnetic hysteresis polarity is insensitive to $E$[10].

The abrupt switching of $R_{xy}(E)$ indicates a switchable electric polarization $P$, while the electric-field-controlled reversal of magnetic hysteresis polarity suggests coupling between $P$ and orbital magnetization $M$. We therefore propose a phenomenological multiferroic free energy of the form $F = -\lambda P \cdot M \cdot B$, where $\lambda$ is a coupling constant. Simulations based on this free-energy term successfully reproduce the key experimental features (see Fig. S15).

This free-energy form is distinct from that of the previously reported multiferroic orbital magnetism near $E = 0$, which is also observed in our system in the triangular region marked by the green circle in Fig. 1e. In that case, Ref. 10 describes the free energy as $F = -M \cdot B = -(\alpha V E) \cdot B$, where $M = \alpha V E$ is the orbital magnetization, $V$ denotes valley polarization, $\alpha$ is a constant. Because the underlying order parameter is valley polarization and no spontaneous electric polarization exists at $E = 0$, we refer to that state as ferrovalley orbital magnetism to distinguish it from the ferroelectric orbital magnetism uncovered here. In the ferrovalley case, the orbital moment remains aligned with the external magnetic field, and the electric field only tunes the magnitude of the orbital magnetization and the hysteresis loop size.

The microscopic origin of the hysteretic electric dipole remains unclear. One possible scenario is a 3:1 isospin-polarized configuration in which three positive (negative) emergent gaps coexist with one negative (positive) gap among the four isospin flavors. This interpretation is consistent with the phase's location between the fully layer-polarized band insulator and the layer-unpolarized correlated insulator in the phase diagram.

# Methods

**Device preparations.** The hexalayer graphene and hBN flakes were prepared by mechanical exfoliation of high-quality graphite and hBN single crystal onto $SiO_2$/Si substrates. $SiO_2$/Si substrates are oxygen-plasma treated before graphene exfoliation to elevate the yield of rhombohedral graphene layers. The rhombohedral domains were identified by using Raman spectroscopy with a 532 nm laser at room temperature[33] and then isolated by anodic oxidation lithography with atomic force microscope (see Fig. S16)[34]. Our heterostructure was assembled with a dry transfer technique[35] using poly(bisphenol A carbonate)/polydimethylsiloxane stamp. We use the following sequence for pickup: graphite (top gate)/hBN/rhombohedral graphene/hBN/graphite (bottom gate). After assembling all the layers, we release the stacks on $SiO_2$/Si chips. The geometry of the Hall bar devices was defined by combining $CHF_3/O_2$ and $O_2$ etching. After Hall bar etching, we deposit Cr(5 nm)/Pd(15 nm)/Au(80 nm) as contacts to the rhombohedral graphene[35]. For better contact, we performed another etching by $CHF_3/O_2$ plasma right before metal evaporation. Three devices were fabricated and measured with similar results.

**Transport measurements.** The devices were studied in an Oxford Proteox dilution refrigerator with a base temperature of 9 mK. Stanford SR860 lock-in amplifiers were used for DC transport measurements with the frequency ranging from 17.777 to 37.777 Hz. We measure longitudinal four-terminal voltages, bias currents, and Hall voltages simultaneously to obtain $R_{xx}$ and $R_{xy}$. For current dependent $dV/dI$ measurements, we use a Keithley 2450 with a 4.7 MΩ serial resistance to generate DC current bias. For better electrical contact, we apply voltages to the global silicon gate ranging from -90 to -100 V in the hole-doping side, and 95 V for the electron side.

**Tight-binding calculations.** The single-particle band structure of rhombohedral hexalayer graphene was calculated using a tight-binding model for the 12-band continuum Hamiltonian, with parameters taken from previous literature[4,36].

In brief, we construct the Hamiltonian:

$$H_{6lRG}(\mathbf{k}) = \begin{bmatrix} h_l^{(0)} & h^{(1)} & h^{(2)} & 0 & 0 & 0 \\ (h^{(1)})^\dagger & h_l^{(0)} & h^{(1)} & h^{(2)} & 0 & 0 \\ (h^{(2)})^\dagger & (h^{(1)})^\dagger & h_l^{(0)} & h^{(1)} & h^{(2)} & 0 \\ 0 & (h^{(2)})^\dagger & (h^{(1)})^\dagger & h_l^{(0)} & h^{(1)} & h^{(2)} \\ 0 & 0 & (h^{(2)})^\dagger & (h^{(1)})^\dagger & h_l^{(0)} & h^{(1)} \\ 0 & 0 & 0 & (h^{(2)})^\dagger & (h^{(1)})^\dagger & h_l^{(0)} \end{bmatrix}$$

which is a matrix in layer space with entries in sublattice space,

$$h_l^{(0)}(\mathbf{k}) = \begin{pmatrix} u_{Al} & -t_0 f_{\mathbf{k}} \\ -t_0 f_{\mathbf{k}}^* & u_{Bl} \end{pmatrix}$$

$$h^{(1)}(\mathbf{k}) = \begin{pmatrix} t_4 f_{\mathbf{k}} & t_3 f_{\mathbf{k}}^* \\ t_1 & t_4 f_{\mathbf{k}} \end{pmatrix}$$

$$h^{(2)}(\mathbf{k}) = \begin{pmatrix} 0 & t_2/2 \\ 0 & 0 \end{pmatrix}$$

$$f_{\mathbf{k}} = \sum_{i=0}^{2} e^{i\mathbf{k}\cdot\boldsymbol{\delta}_i}$$

In the numerical simulation, we use phenomenological parameters,

$$u_{\sigma l} = (7 - N_l)U$$

$$(t_0, t_1, t_2, t_3, t_4) = (3120,\ 377,\ -20.6,\ 290,\ 120)\ \text{meV}$$

where we use the $u_{\sigma l}$ as the onsite potential, $N_l$ is the layer index. Parameters $t_1$, $t_3$, and $t_4$ quantify the hopping between atoms on adjacent carbon layers, and parameter $t_0$ corresponds to the hopping between nearest neighbor atoms, specifically the A and B sublattices in a given layer. In addition, $t_2$ in rhombohedral graphene corresponds to hopping between $A_1$ and $B_3$ atoms, which are located on the outer layers and are on top of each other.

The Hamiltonian eigenvalues were computed using standard diagonalization routines with a uniform distribution in the Brillouin zone. The DOS was obtained by constructing energy histograms from all eigenvalues across the sampled $k$-points, with energies binned into intervals. The final DOS was normalized by the total number of states (number of $k$-points × 12 bands) to obtain the density per unit energy per unit cell.

An interlayer potential difference was modelled as a layer-dependent on-site energy shift $\delta$, varied to simulate the effect of a perpendicular electric field. For each $\delta$, the Hamiltonian was diagonalized on a uniform triangular grid of $k$-points in the irreducible Brillouin zone, and the density of states was computed from the histogram of all eigenvalues. Repeating this procedure over a range of $\delta$ produced the two-dimensional map DOS(E, $\delta$). Layer polarization shown in Fig. 1b is defined as $|\psi_{\text{top}}|^2 - |\psi_{bot}|^2/|\psi_{\text{top}}|^2 + |\psi_{bot}|^2$.

**Simulation of ferroelectric orbital magnetism.** To capture the coupled ferroelectric and orbital magnetic order observed experimentally, we adopt a phenomenological Landau free-energy functional of the form

$$F = -\frac{a}{2}P^2 + \frac{b}{4}P^4 - \frac{\alpha}{2}M^2 + \frac{\beta}{4}M^2 - E \cdot P - B \cdot M - \lambda P \cdot M \cdot B$$

Here $P$ denotes the electric polarization, $M$ the orbital magnetization, $E$ the applied electric field, and $B$ the perpendicular magnetic field. The first two (second two) terms describe the conventional double-well potential associated with spontaneous polarization (magnetization). The bilinear coupling term proportional to $\lambda$ captures the magnetoelectric interaction responsible for the multiferroic response.

For the simulations shown in Fig. S15, we use dimensionless parameters:

$$(a, b, \alpha, \beta, \lambda) = (1, 1, 1, 1, 2.5).$$

The equilibrium states are obtained numerically by minimizing $F(P, M)$ with respect to $P$ and $M$ at fixed ($E, B$). To reproduce hysteretic behavior under field sweeps, the minimization at each field point is initialized from the converged solution at the preceding ($E, B$) value, thereby tracking local free-energy minima rather than enforcing global minimization.

**Extraction of effective mass from Fermiology.** To study the origin of the temperature sensitive behavior, we measured the temperature dependent quantum oscillation amplitude of different Fermi surfaces at SC1 region and extracted the effective mass from modified Lifshitz-Kosevich formula $\frac{\Delta R}{R_a} = \frac{\lambda^*}{sin\ (\lambda^*)}$, where $\Delta R$ and $R_a$ are the oscillation amplitude in FFT result and amplitude constant, and $\lambda^* = \frac{2\pi^2 k_B T m^*}{\hbar e B_{eff}}$ is the damping factor where $k_B$, $T$, $B_{eff}$, $\hbar$, $e$, and $m^*$ are Boltzmann constant, temperature, effective magnetic field, Plank constant, and elementary charge, and carrier effective mass (see details in Fig. S7)[37]. Here we determined the effective field $B_{eff} = 2/(1/B_{min} + 1/B_{max})$, where $B_{min}$ and $B_{max}$ are 0.5 and 1.6 T, respectively, corresponding to the lower and upper magnetic field limit we used for collecting raw quantum oscillation data for the FFT results. Since the FFT result always has finite background signal which becomes significant at low frequency, we added a background constant in this formula as a fitting parameter.

## Acknowledgments

We acknowledge helpful discussions with C. Chang, J. Zhu, T. Cao, M. Yankowitz, C. Lewandowski, C. Huang, D. Chowdhury, and C. Jian. X. L. acknowledges support from the National Science Foundation through the CAREER program under Award No. DMR-2442363. J. D. acknowledges support from the New Frontier Grant, College of Arts & Sciences, Cornell University. The device fabrication is performed in part in Cornell Center for Materials Research and in part at the Cornell NanoScale Facility, a member of the National Nanotechnology Coordinated Infrastructure (NNCI), which is supported by the National Science Foundation (Grant NNCI-2025233). K.W. and T.T. acknowledge support from the JSPS KAKENHI (Grant Numbers 21H05233 and 23H02052), the CREST (JPMJCR24A5), JST and World Premier International Research Center Initiative (WPI), MEXT, Japan.

## Author Contributions

J. D., J. X., H. L., and X. L. conceived the project. J. D., J. X., and H. L. fabricated the device. J. D. performed the transport measurement and analyzed the data with the help of H. L. and J. X. K.W. and T. T. provided hBN crystals. X. L. supervised the project and provided input on data analysis. J. D. and X.L. wrote the manuscript with input from all authors.

## Competing interests

The authors declare no competing interests.

## Additional information

Supplementary information is available for this paper at XXX.

**Supplementary Information for:**

# Superconductivity and Ferroelectric Orbital Magnetism in Semimetallic Rhombohedral Hexalayer Graphene

*Jinghao Deng[1#], Jiabin Xie[1#], Hongyuan Li[1#], Takashi Taniguchi[2], Kenji Watanabe[3], Jie Shan[4,5], Kin Fai Mak[1,5], Xiaomeng Liu[1*]*

*[1]Laboratory of Atomic and Solid State Physics, Cornell University, Ithaca, NY, USA*

*[2]Research Center for Materials Nanoarchitectonics, National Institute for Materials Science, 1-1 Namiki, Tsukuba 305-0044, Japan*

*[3]Research Center for Electronic and Optical Materials, National Institute for Materials Science, 1-1 Namiki, Tsukuba 305-0044, Japan*

*[4]School of Applied and Engineering Physics, Cornell University, Ithaca, NY, USA*

*[5]Max Planck Institute for the Structure and Dynamics of Matter, Hamburg, Germany*

*[#]These authors contributed equally*

**Correspondence author E-mail: xl956@cornell.edu*

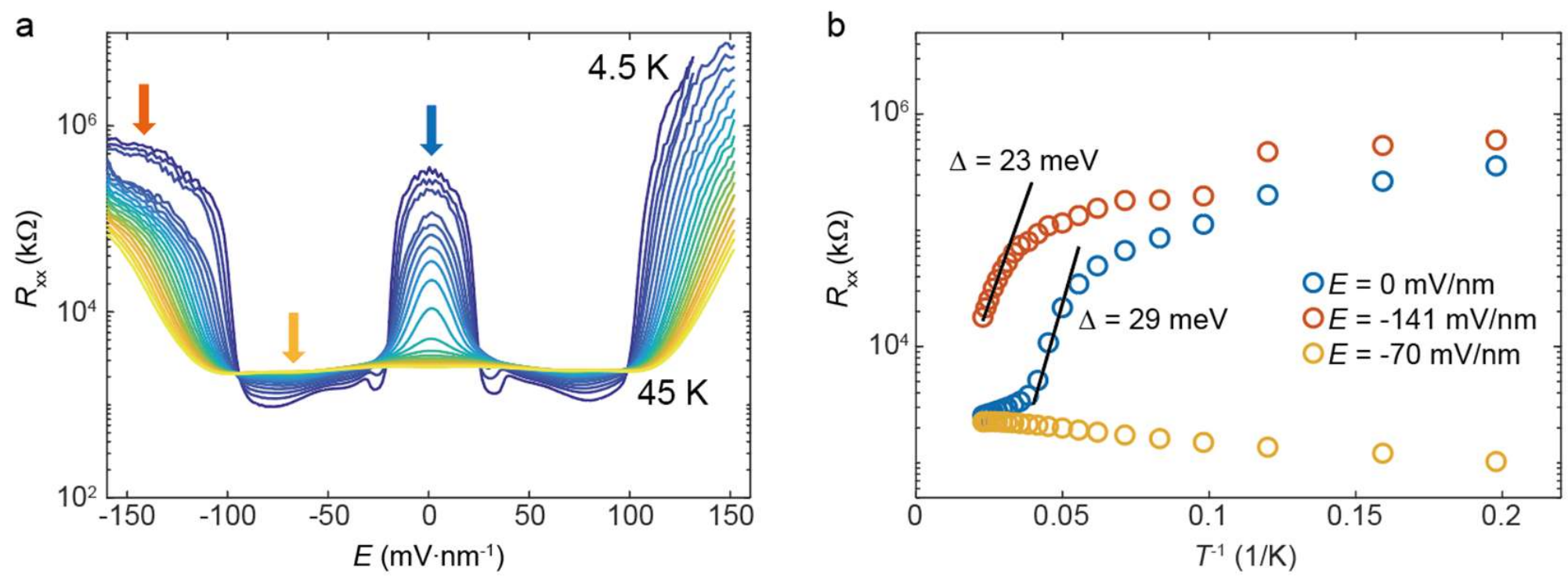

**Figure S1 | Temperature-dependent correlated and band insulating states. a**, Temperature-dependent $R_{xx}$ measured at $n = 0$ from 4.5 K to 45 K. **b,** $R_{xx}$ versus $1/T$ plot at different $E$ extracted from the corresponding arrows in **a**. At zero $E$ (blue) and high $E$ (orange) regions, the $R_{xx}$ increases with decreasing temperature, exhibiting insulating behavior at low temperatures. Between these two insulating phases (yellow), the $R_{xx}$ decreases with decreasing temperature, indicating typical metallic behavior. We extract the transport gap $\Delta$ by using the formula $R_{xx} \propto e^{-\Delta/2k_BT}$, as shown by the black line in **b**. At high temperatures ($T > 28$ K) and zero electric field, the slope dramatically decreases, indicating that the gap value decreases with increasing temperature. This observation is consistent with the previously observed correlated insulating states in tetralayer and pentalayer graphene[1,2].

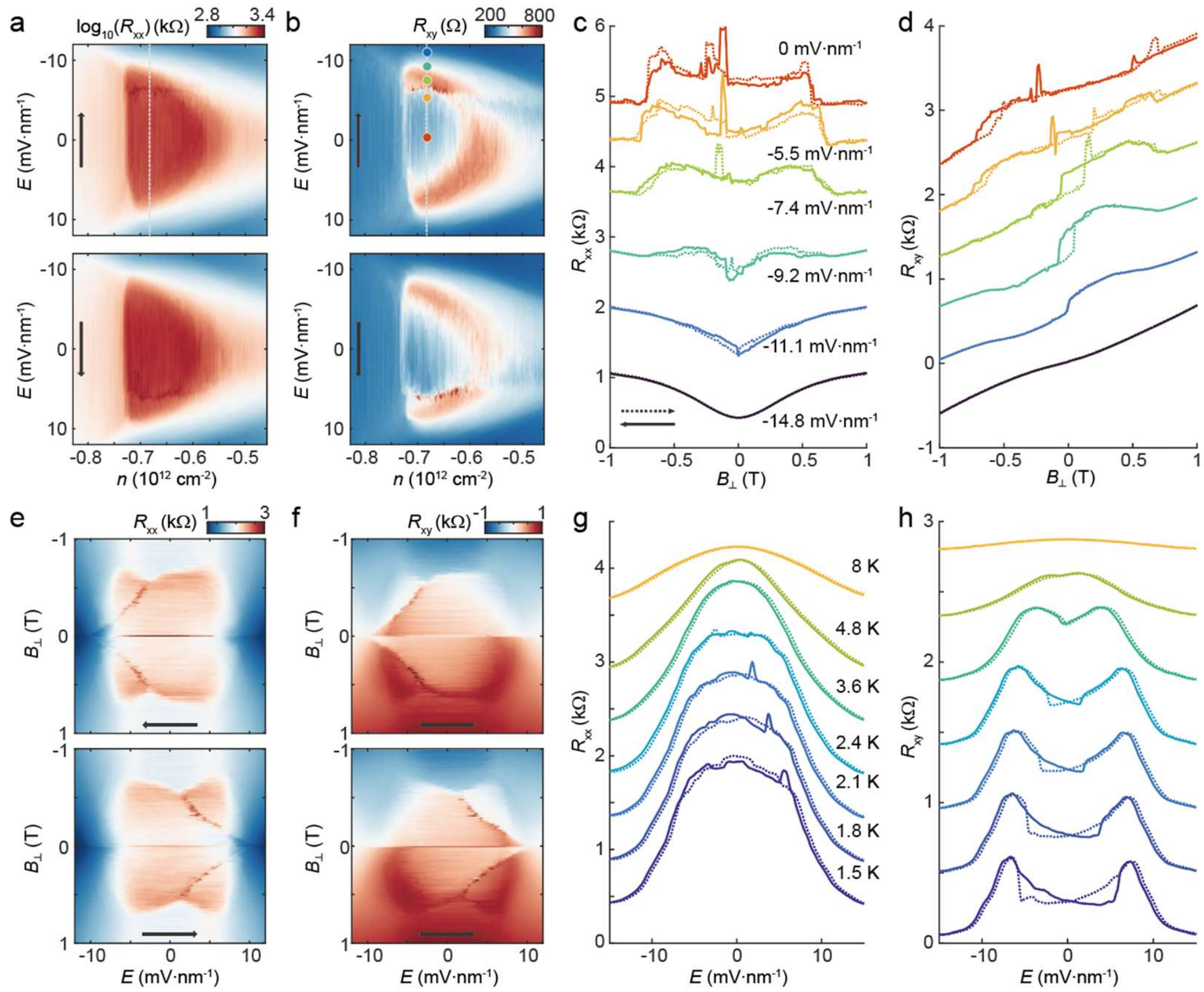

**Figure S2 | Electric field and magnetic field control of ferrovalley orbital ferromagnetic states. a, b,** $n$ and $E$ dependent $R_{xx}$ (**a**) and $R_{xy}$ (**b**), the upper and lower panels represent results with forward and backward scanning of $E$, respectively. A small magnetic field $B_{\perp}$ = 50 mT is added to stabilize the magnetic order. Abrupt changes in $R_{xy}$ at around critical electric field $E_c$ = -6 mV·nm$^{-1}$ (forward) and 6 mV·nm$^{-1}$ (backward) indicate that the magnetic order can be reversed by the electric field. **c, d,** Electric-field-dependent magnetic hysteresis of $R_{xx}$ (**c**) and $R_{xy}$ (**d**) recorded at $n$ = -0.68 × 10$^{12}$·cm$^{-2}$. The location of each curve in $n$-$E$ parameter space is marked by dots in **b** with corresponding color, except for the black ones at $E$ = -14.8 mV·nm$^{-1}$, which are out of the range in **b**. Dashed and solid curves represent the forward and backward scanning of $B_{\perp}$. The hysteresis behavior in $R_{xy}$ reveals the existence of orbital magnetization, which can be further tuned by the electric field due to the tunable Berry curvature. **e, f,** $B_{\perp}$ and $E$ dependent $R_{xx}$ (**e**) and $R_{xy}$ (**f**) at $n$ = -0.68 × 10$^{12}$·cm$^{-2}$, the upper and lower panels represent results with forward and backward scanning of $E$, respectively. The $B_{\perp}$ dependent $2\Delta E_c$ indicates the coupling between $E$ and magnetic ordering. **g, h,** Temperature-dependent $E$ hysteresis of $R_{xx}$ (**g**) and $R_{xy}$ (**h**) at $n$ = -0.68 × 10$^{12}$·cm$^{-2}$. The butterfly-like hysteresis behavior is weakened with elevated temperature up to 3.6 K and eventually vanishes at 4.8 K, suggesting a multiferroic transition temperature $T$ > 3.6 K. All data in this figure are acquired at 1 K except for the temperature-dependent measurement.

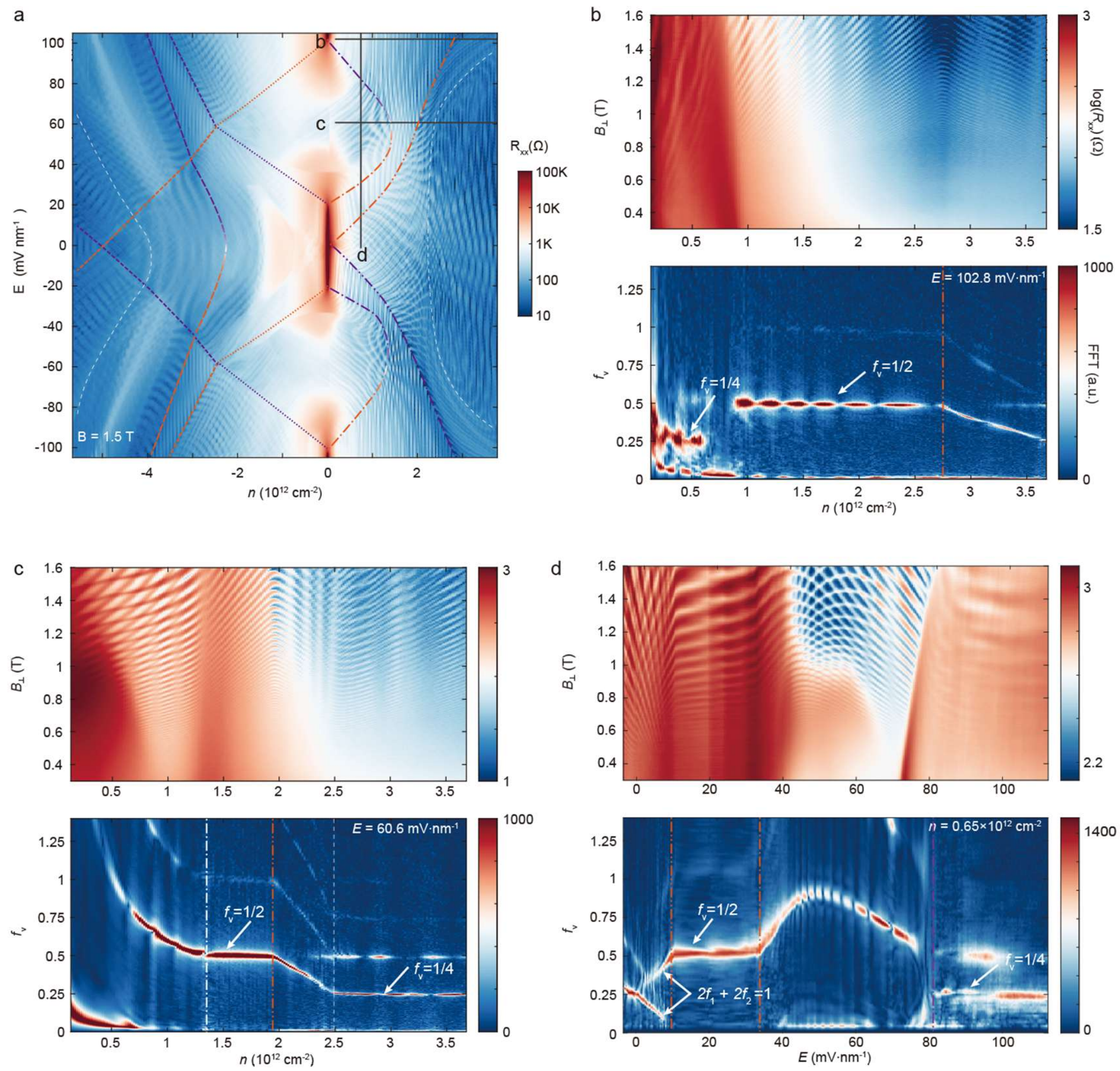

**Figure S3 | Fermiology at electron-doped side. a**, Longitudinal resistance vs *n* and *E* at 1.5 T (same as Fig. 2a). The phase boundaries as shown in **a** are also marked in **b**, **c**, and **d**. **b, c,** and **d,** Quantum oscillations and their FFT results acquired at $E$ = 102.8, 60.6 mV·nm$^{-1}$, and $n$ = -0.65 × 10$^{12}$·cm$^{-2}$, respectively. The location of the linecuts is marked by black solid lines in **a**.

In **b**, we show that doping the band insulator will first lead to a Fermi surface with a four-fold degeneracy, which can be explained by either a simple full metal or a two-fold degenerate nematic half metal, which were observed at low carrier density area in trilayer rhombohedral graphene[3]. Increasing the doping level can tune the system into a half-metal state with a two-fold degeneracy and then transfer to a partially isospin (flavor) polarized state with a violation of the sum rule. As shown in Fig. 2g, this partially flavor polarized stems from the involvement of the conduction band bottom in flavor B. In **c**, we show a quantum oscillation result of doping the band overlapped semimetallic regime located between the band insulator and correlated insulator. Here, the semimetallic regime shows partially isospin polarized features, suggesting the complex Fermi surface nature in the band overlapped semimetal region, and further transfer to a half metal state with the increasing doping level. Keeping increasing the carrier density can lead to a half-metal to full metal transition with a partially isospin polarized transition state. In **d**, constant carrier density quantum oscillation results show a clear *E-dependent* band evolution. At zero electric field, the Fermi surface is formed by a four-fold degenerate full metal state. Increasing the electric field elevates the conduction band edge energy in one of the flavors (*i.e.*, flavor B as we showed in Fig. 2), leading to an immediate splitting of the four-fold degenerated band into a majority and a minority two-fold degenerated bands. When $E > 10$ mV·nm$^{-1}$, an electric field triggered Lifshitz transition appears, where the minority Fermi surface disappears and transitions to a half-metal state. Keep increasing carrier density tunes the system into the partially isospin polarized and four-fold degenerated metallic states as we discussed in **b** and **c**. We noticed intrinsic segmented features in both raw quantum oscillation data and their FFT results in all three figures. These features are commonly observed in multiband system, where two sets of quantum oscillations from nearby Fermi surfaces interference constructively or destructively, forming complicated pattern[4,5].

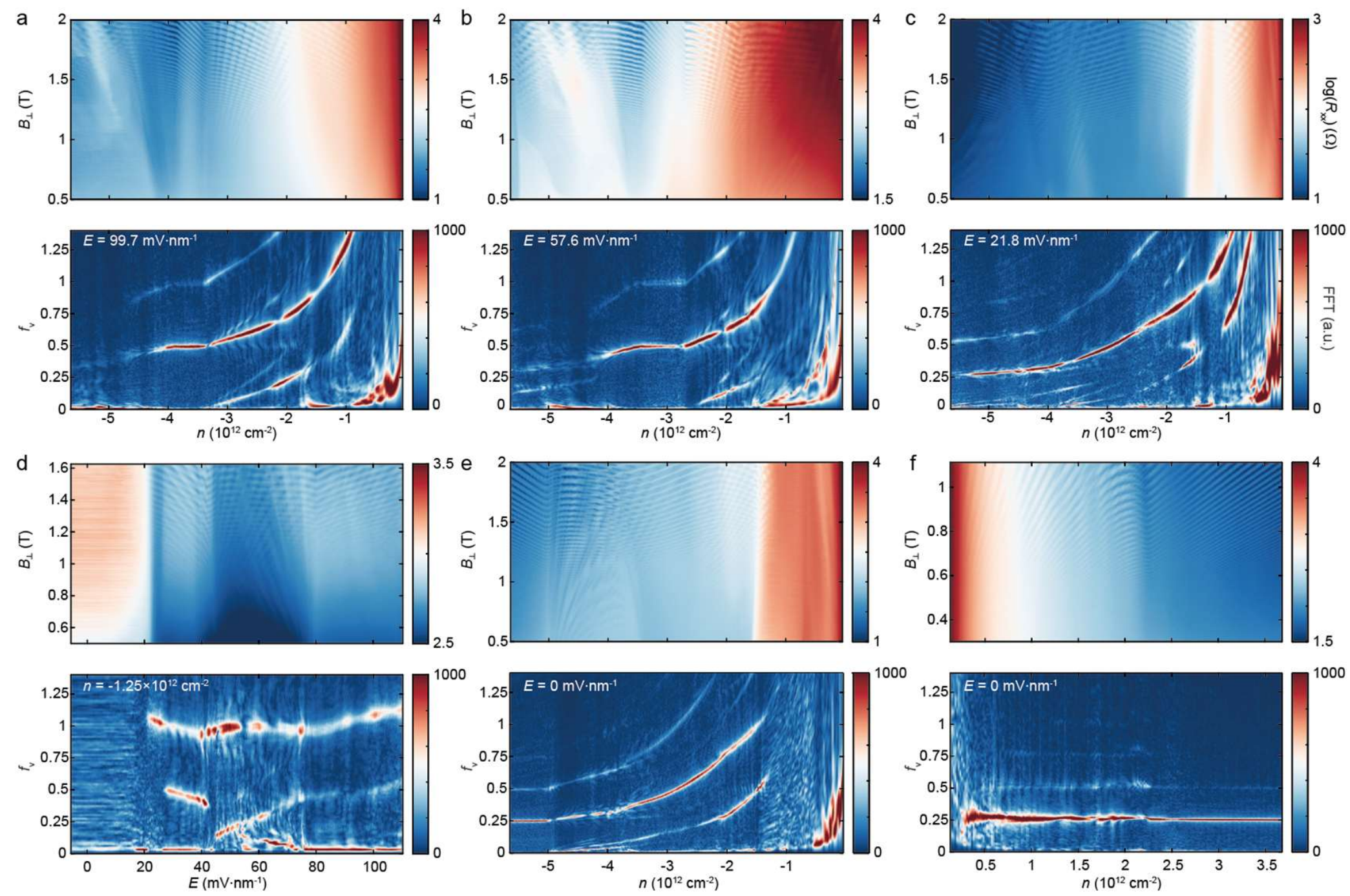

**Figure S4 | Raw data of the quantum oscillation and the corresponding FFT results. a to f**, Quantum oscillation and their FFT results as shown in Fig. 2. **a, b, c,** and **d** correspond to the Figs. 2b, c, d, and f, respectively. **e** and **f** correspond to the hole-doped and electron-doped region in Fig. 2e

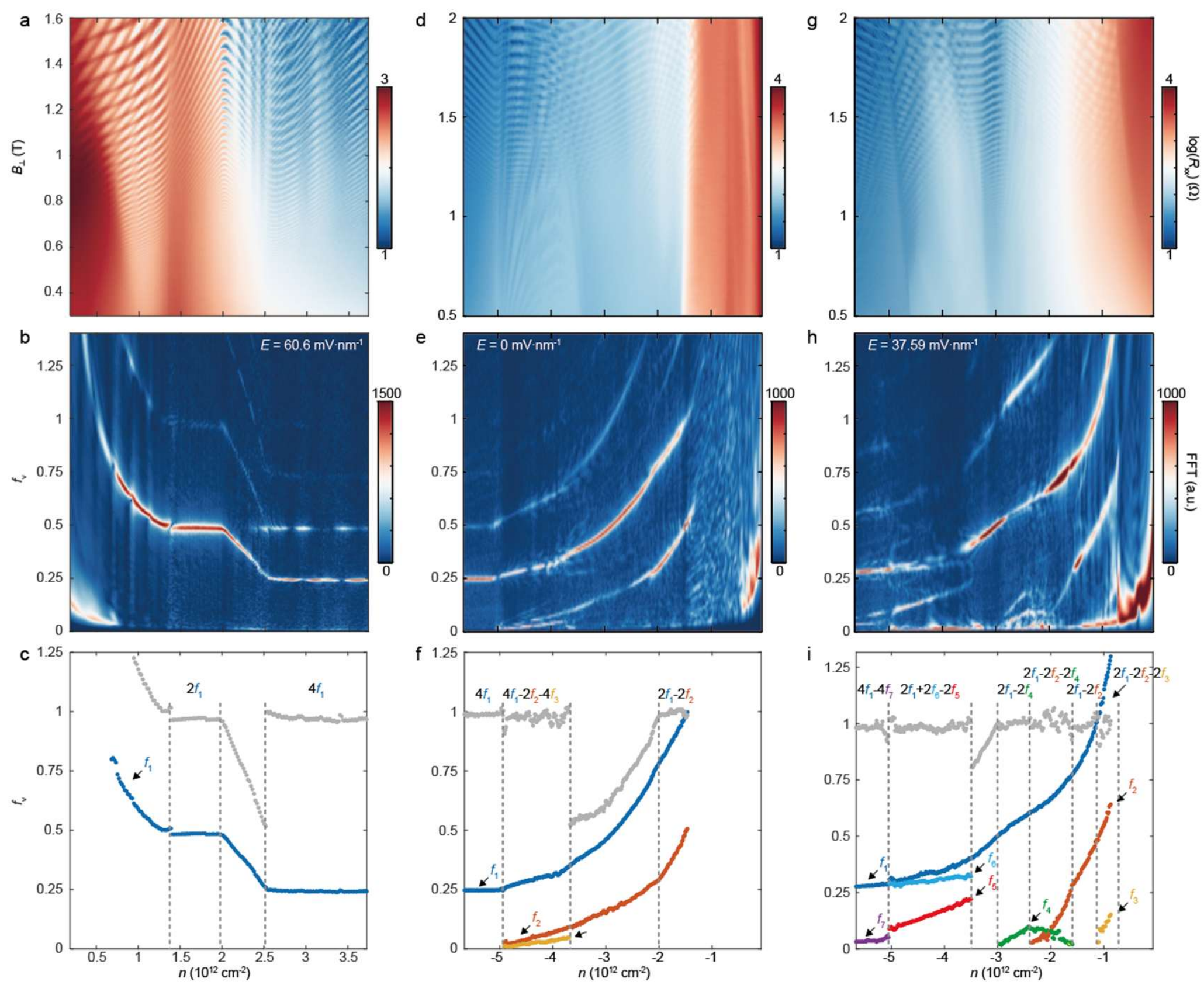

**Figure S5 | Sum rule discussion of quantum oscillation results and their violation at partially flavor polarized states. a** to **c,** Quantum oscillation (**a**), FFT results (**b**), and extracted frequency peaks and their sum rule illustration (**c**) performed at $E$ = 60.6 mV·nm$^{-1}$ linecut at electron side. **d** to **f,** Quantum oscillation (**d**), FFT results (**e**), and extracted frequency peaks and their sum rule illustration (**f**) performed at $E$ = 0 mV·nm$^{-1}$ linecut at hole side. **g** to **i,** Quantum oscillation (**g**), FFT results (**h**), and extracted frequency peaks and their sum rule illustration (**i**) performed at $E$ = 37.59 mV·nm$^{-1}$ linecut. In extracted results **c**, **f**, and **i**, the extracted frequency peaks are colorfully marked, and their sum rule results are calculated and shown in gray dots.

The sum for normalized frequency ($f_\nu$) should, in principle, follow the sum rule $\sum_i s_i k_i f_i = 1$, where $s$ can be either -1 or 1, representing electron- or hole-like Fermi surface, and $k$ indicates the degeneracy of the Fermi surface. In **c**, we show two representative half-metal and full metal phases on the electron-doped side with a simple two-fold ($2f_1 = 1$) and four-fold degenerated ($4f_1 = 1$) enclosed Fermi surface. However, we noticed some areas showing a violation of sum rules in connection with these phases. These phases were commonly observed in other rhombohedral multilayer graphene system[6,7], which is usually observed between symmetry-breaking phases and therefore attributed to partially flavor polarization. A possible explanation of these violations is that these transition phases always involve multiple small Fermi surfaces with high degeneracy and high effective mass[8], which is difficult to resolve by quantum oscillation measurements. Another possible scenario involves a Fermi-surface reconstruction that suppresses quantum oscillations. For example, Ref. (arXiv:2505.05414) proposed a crescent-shaped Fermi surface, in which a strong asymmetry between forward- and backward-moving carriers eliminates quantum oscillations.

In **f**, we discuss the Fermi surface evolution versus carrier density at the $E = 0$ linecut at hole-doped side. Upon doping the correlated insulator, the system transitions into the ferrovalley orbital magnetic phase, where quantum oscillations are absent. Recent theoretical work suggests that the absence of quantum oscillations may originate from the distinctive Fermi-surface geometry in the orbital magnetic regime of multilayer rhombohedral graphene[9]. On the left of the orbital magnetic region, the quantum oscillation signal appears with the sum rule $2f_1 - 2f_2 = 1$, indicating a typical half-metallic hole-doped annular Fermi surface. Increasing the hole density bring the system across a partially flavor polarized state that violates the sum rule and then into a complex multi-Fermi surface region, before reaching the simple full metallic state ($4f_1 = 1$). In the complex multi-Fermi surface region, we found a sum rule of $4f_1 - 2f_2 - 4f_3 = 1$ that may points to a possible nematic ordering[3].

In **i**, we discuss the Fermi surface evolution versus carrier density at $E = 37.59$ mV·nm$^{-1}$ linecut across the SC1 state at hole-doped side. At low carrier density $|n| < 1.6 \times 10^{12}$·cm$^{-2}$, the quantum oscillation sum rule can be understood from a single particle picture as we illustrated in Fig. 2i. The system starts with an electron-hole overlapped semimetallic region indicated by sum rule $2f_1 - 2f_2 - 2f_3 = 1$, which can be explained by a combination of majority annular hole-like Fermi surface ($2f_1$ - $2f_2$) with a minority electron-like enclosed Fermi surface ($2f_3$). With the increasing hole doping, the minority electron-like pocket keeps getting smaller and eventually disappears, turning the system into a half-metallic state with a two-fold degenerate annular Fermi surface ($2f_1 - 2f_2 = 1$).

With the increasing hole doping, the emergent electron Fermi surface appears with a two-fold degeneracy, and the whole Fermi surface can be described by the sum rule $2f_1 - 2f_2 - 2f_4 = 1$. With further hole doping, it passes through a partially flavor polarized state before it reaches a region with two additional Fermi surface $f_5$ and $f_6$ with the sum rule $2f_1 + 2f_6 - 2f_5 = 1$ when $|n| > 3.5 \times 10^{12}$·cm$^{-2}$, which is consistent with the band diagram we have shown in Fig. 2i. Here the Fermi level in flavor A has across the bottom of Mexican hat like band structure ($2f_1$) while still keeps in the other flavor B band with an annular Fermi surface ($2f_6 - 2f_5$). When $|n| > 5 \times 10^{12}$·cm$^{-2}$, the Fermi surface transfers into a full metallic annular structure with four-fold degeneracy.

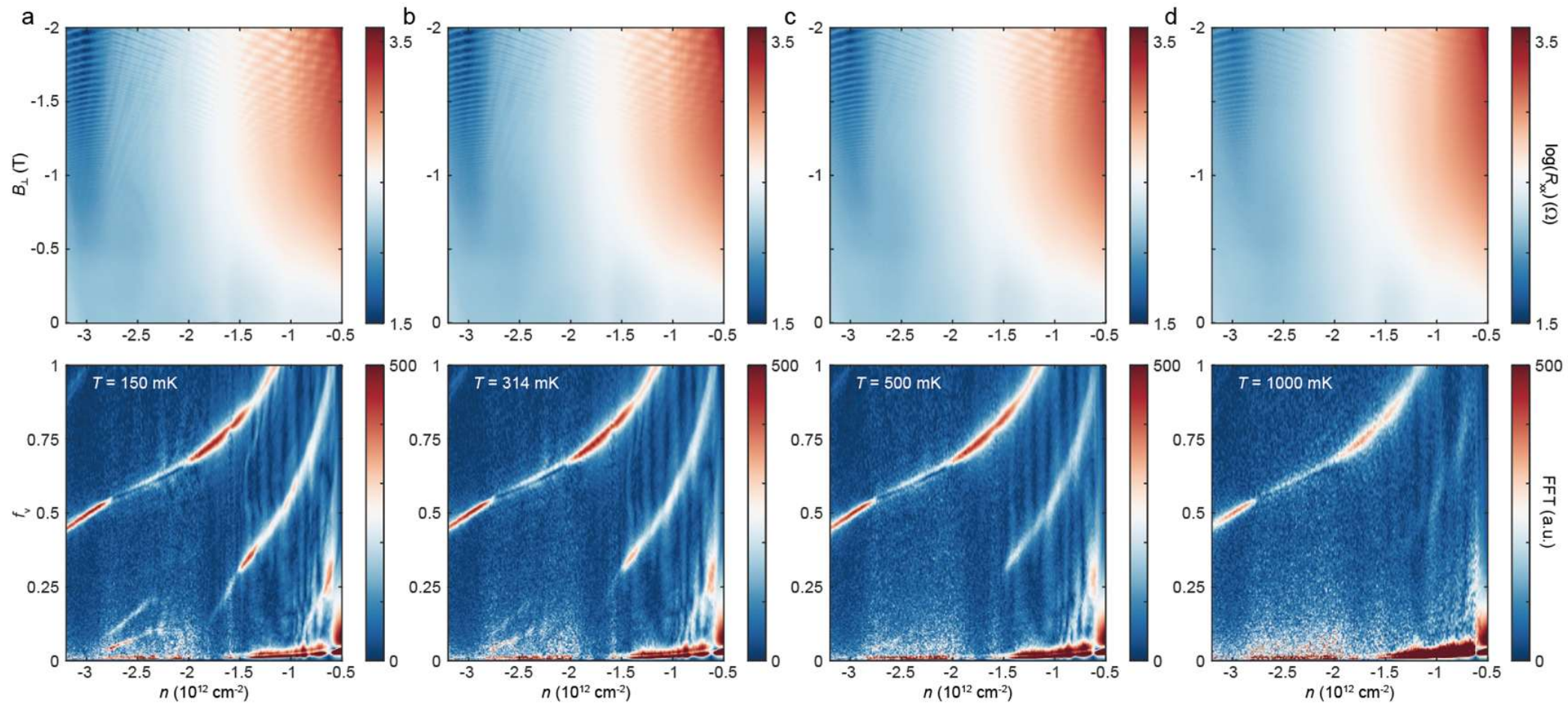

**Figure S6 | Temperature-dependent of the emerging Fermi surfaces FS1 and FS2. a** to **d**, Quantum oscillation and their FFT results performed along the $E$ = 41.6 mV·nm$^{-1}$ linecut at the hole doping region at $T$ = 150, 314, 500, and 1000 mK, respectively. The peak intensity of the emergent Fermi surfaces between $1.5 \times 10^{12}\cdot\text{cm}^{-2} < |n| < 3.3 \times 10^{12}\cdot\text{cm}^{-2}$ decreases and eventually disappears at elevated temperatures.

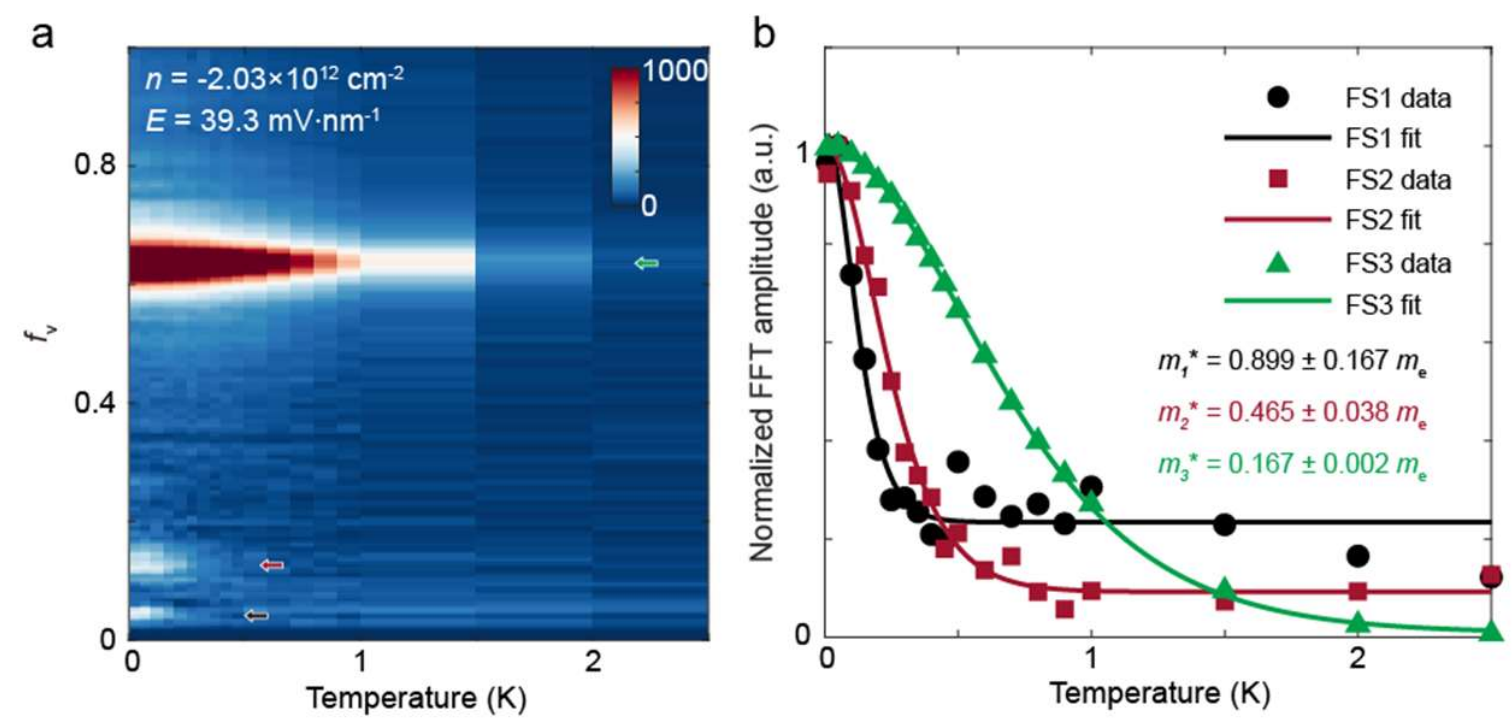

**Figure S7 | Temperature dependence and effective mass extraction of FS1, FS2 and FS3.** **a,** Temperature dependent quantum oscillation measurements at $n$ = -2.03 × $10^{12}$·$cm^{-2}$ and $E$ = 39.3 mV·$nm^{-1}$ in the SC1 region. We observed a dramatic difference of temperature responses for different Fermi surfaces. While the highest frequency peak, which corresponds to FS3, is relatively robust against temperature, the lower frequency peaks, FS1 and FS2, disappear rapidly with increasing temperatures. We extracted the temperature-dependent FFT amplitude normalized with the amplitude at the base temperature for the three Fermi surfaces (marked by corresponding arrows in panel **a**) and plot it in panel **b**. We use modified Lifshitz-Kosevich formula to fit the temperature-dependent amplitude (see Methods) with effective mass as fitting parameter. The fitting results indicate that the emergent FS1 and FS2 Fermi surfaces have elevated effective masses of $m_1^*$ = 0.90 ± 0.17 $m_e$ and $m_2^*$ = 0.46 ± 0.04 $m_e$ while FS3 has the lowest effective mass $m_3^*$ = 0.167 ± 0.002 $m_e$.

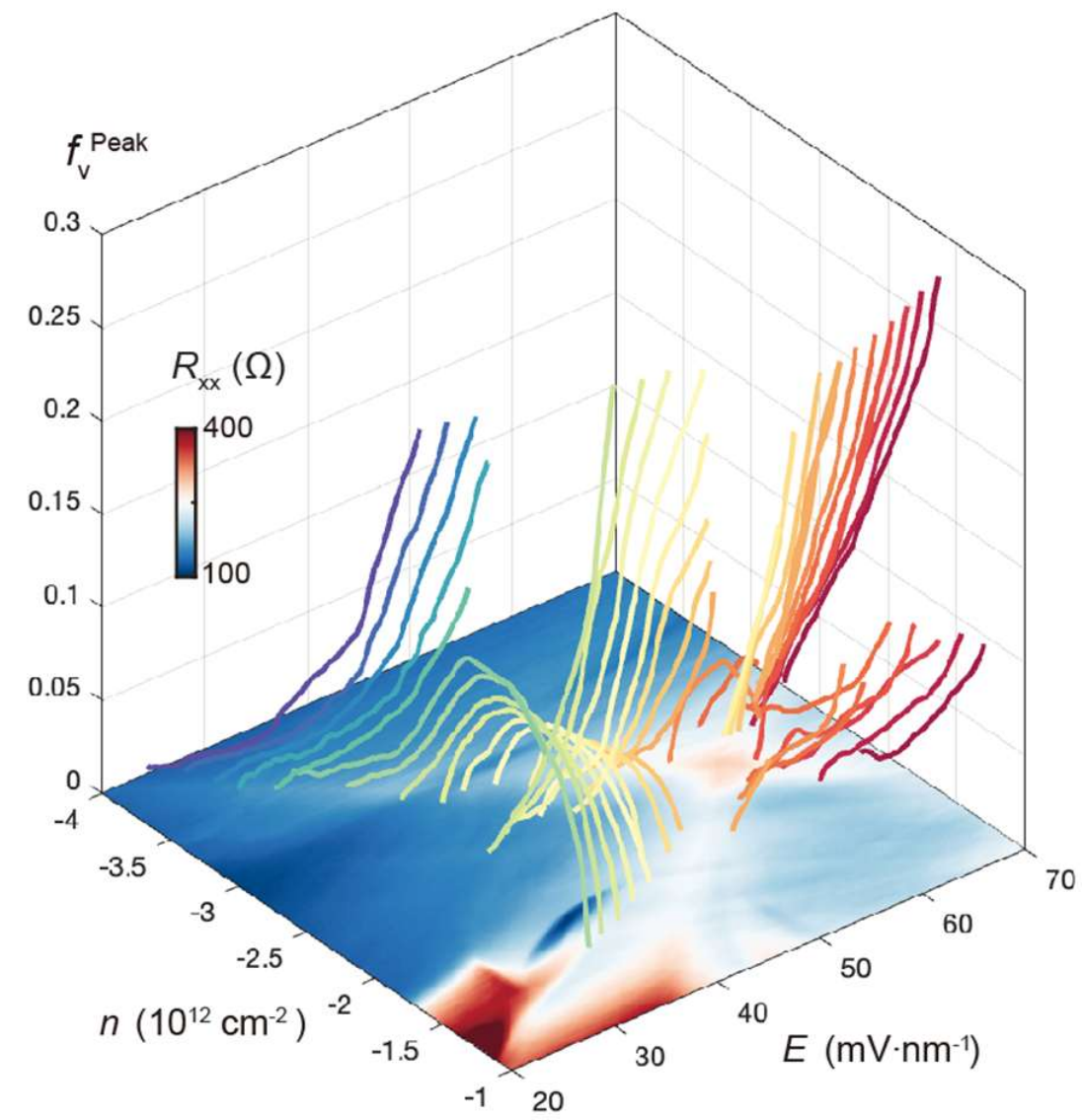

**Figure S8 | 3D evolution of $f_v$ peaks overlaid on the resistance map.** This shows the $f_v$ of two emerging Fermi surfaces, FS1 and FS2, reaches zero at the two boundaries of the leaf-shaped region that hosts superconducting state SC1. Raw quantum oscillations and their FFT results are shown in Figs. S9 and S10.

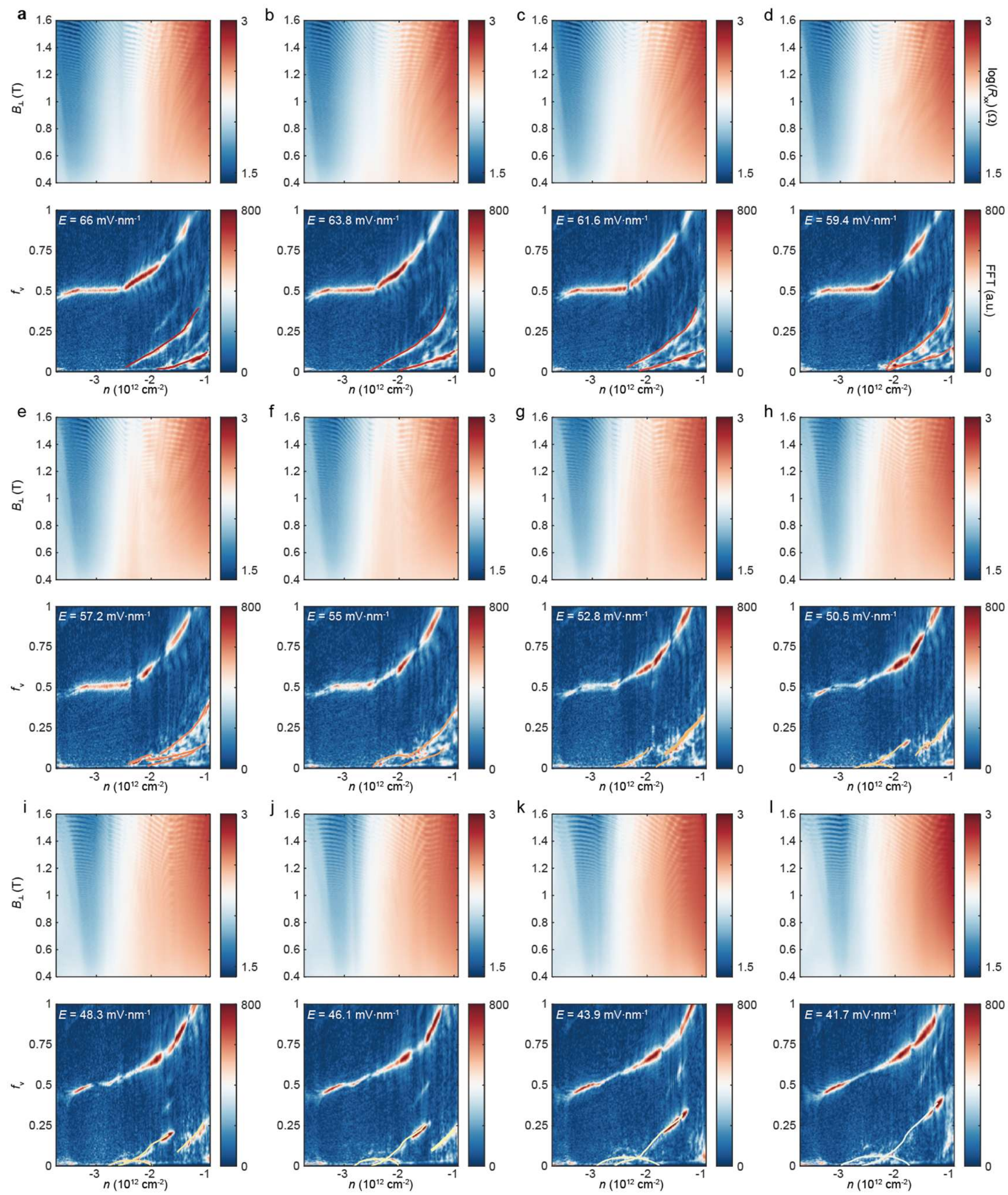

**Figure S9 | Quantum oscillation data near SC1 (part I). a to l,** Raw quantum oscillation data and their FFT results near the SC1 region with $E$ ranging from 66 to 41.7 mV·nm$^{-1}$. The low-frequency peaks that were shown in Fig. S8 are sketched and overlapped on the FFT results by curves with the same color.

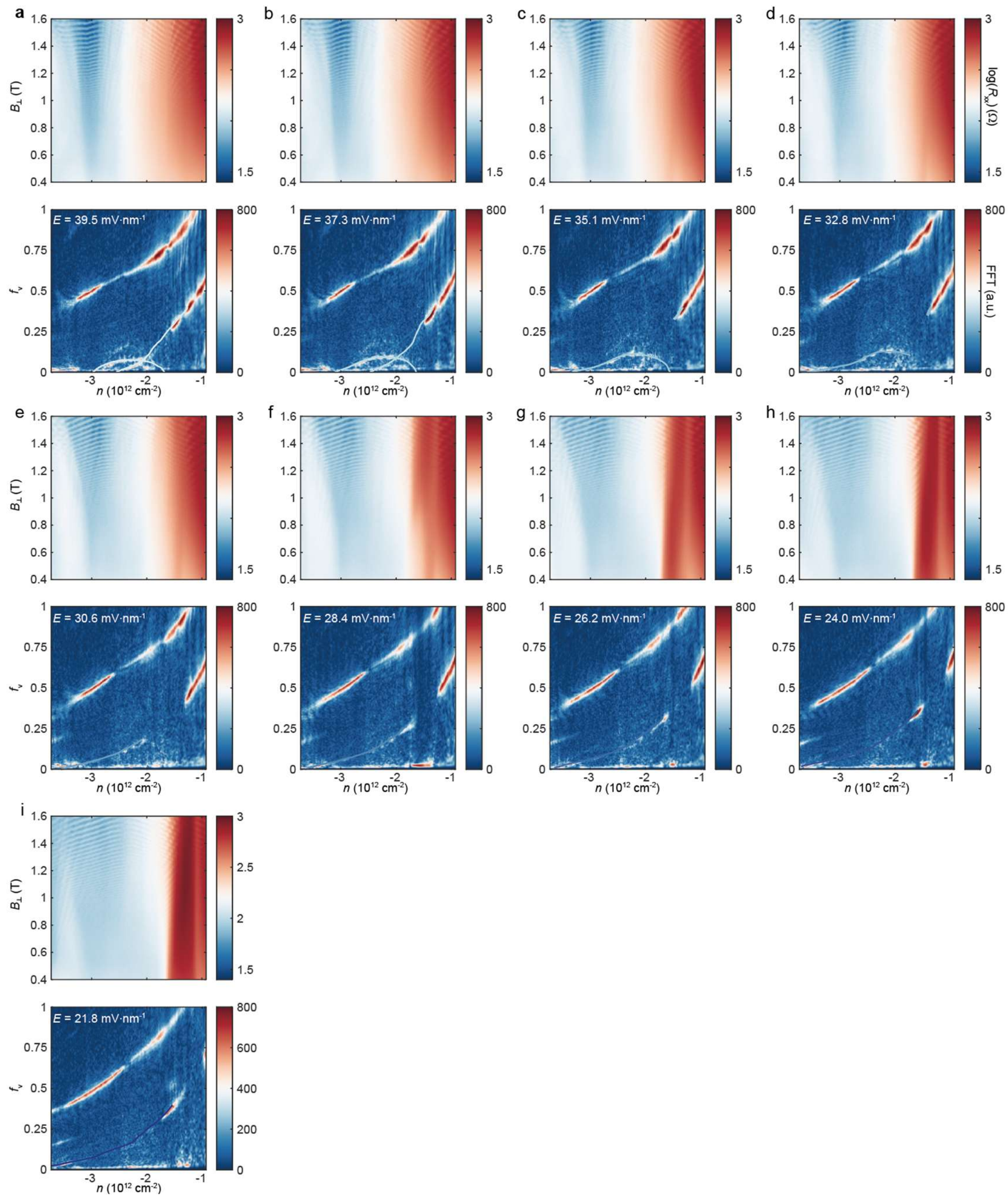

**Figure S10 | Quantum oscillation data near SC1 (part II). a to i,** Raw quantum oscillation data and their FFT results near the SC1 region with $E$ ranging from 39.5 to 21.8 mV·nm$^{-1}$. The low-frequency peaks are sketched by curves with the same color as were shown in Fig. S8.

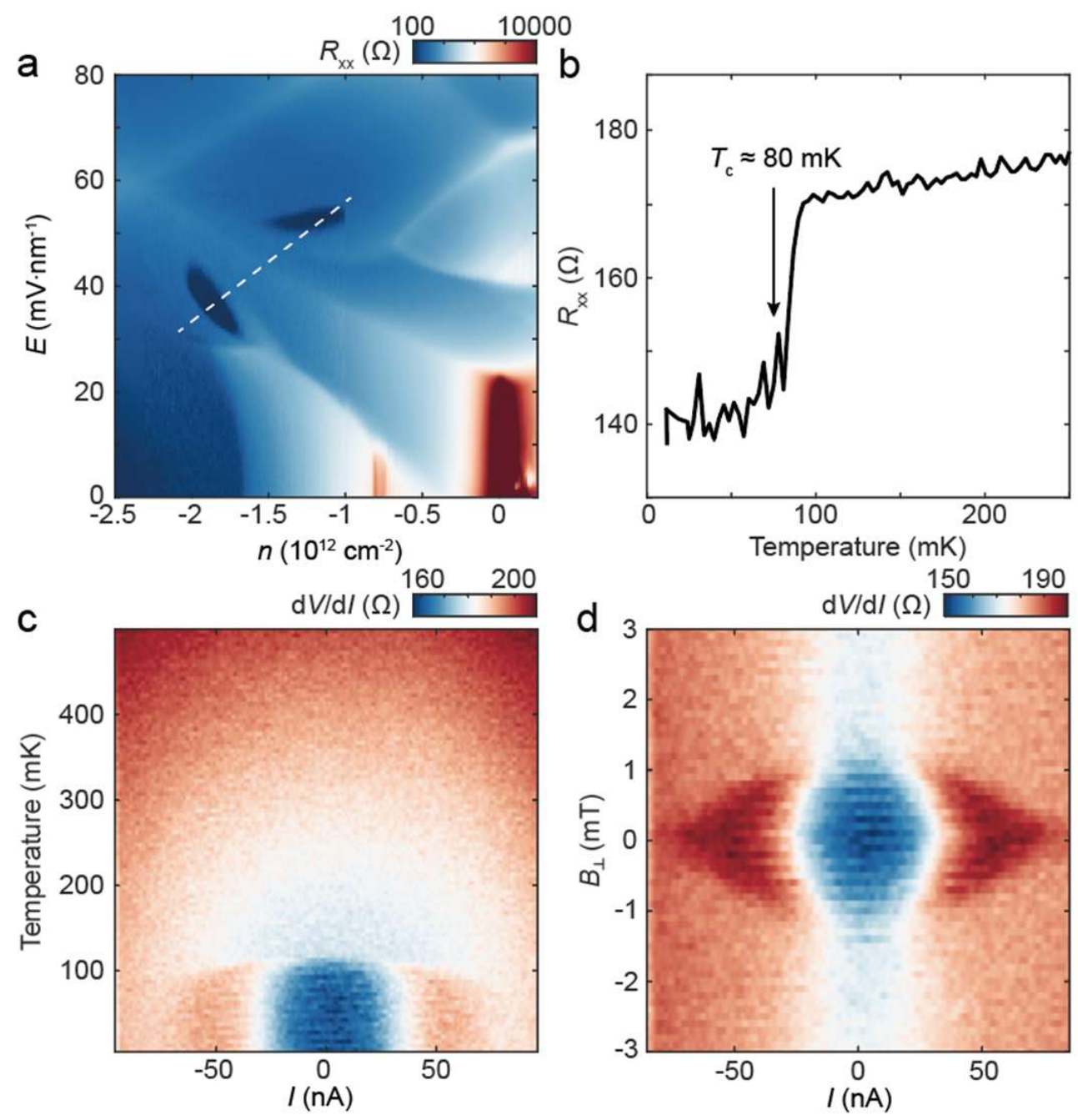

**Figure S11 | Characterization of SC2 and connection between superconductivity and multiferroic. a,** Zoomed-in longitudinal resistance vs *n* and *E*. The white dashed line indicates the location of the linecut shown in Fig. 3g. This plot also demonstrates the proximity of SC1 and SC2 with ferrovalley and ferroelectric orbital magnetism in the phase space. **b,** Temperature-dependent longitudinal resistance of SC2 states. A dramatic decrease in resistance is observed when the temperature drop below $T_c$ ~ 80 mK. **c, d,** Temperature and out-of-plane magnetic field dependent d*V*/d*I* measurements of SC2. The disappearance of non-linear d*V*/d*I* features reveals a similar transition temperature as we showed in **b**, and a critical out-of-plane magnetic field of ~ 1mT.

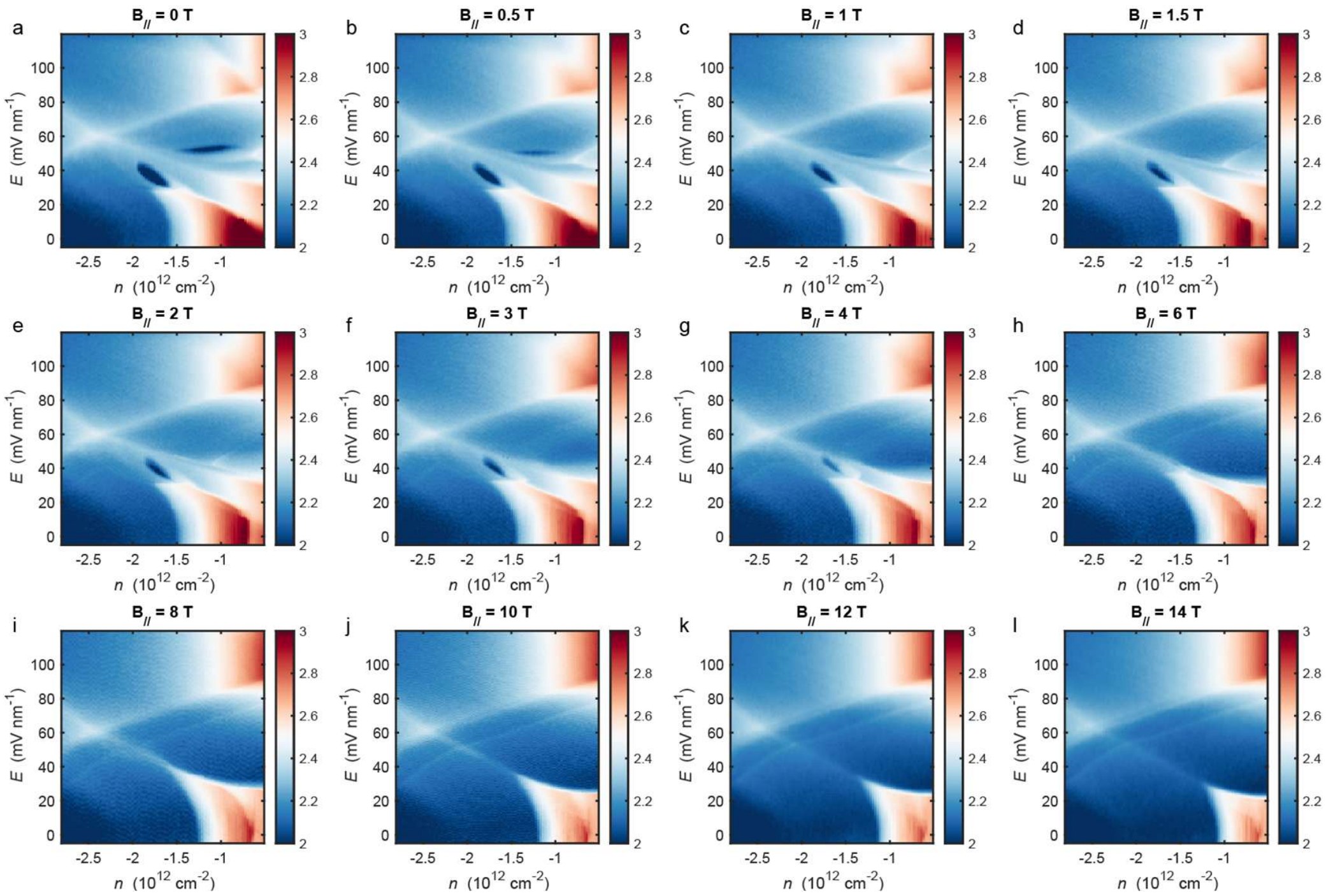

**Figure S12 | Longitudinal resistance maps under various in-plane magnetic fields. a** to **l**, $R_{xx}$ vs $n$ and $E$ under different in-plane magnetic fields from 0 to 14 T. The increasing in-plane magnetic fields not only suppresses the superconducting states but also drastically alter some phase boundaries. Notably, the leaf-shaped region with emergent Fermi surfaces shrinks and become completed eliminated when in-plane field exceed 6T. This observation suggests that the disappearance of the emerging electron-hole Fermi liquid, rather than orbital depairing from a residual out-of-plane field, is the primary mechanism responsible for the observed upper critical in-plane field.

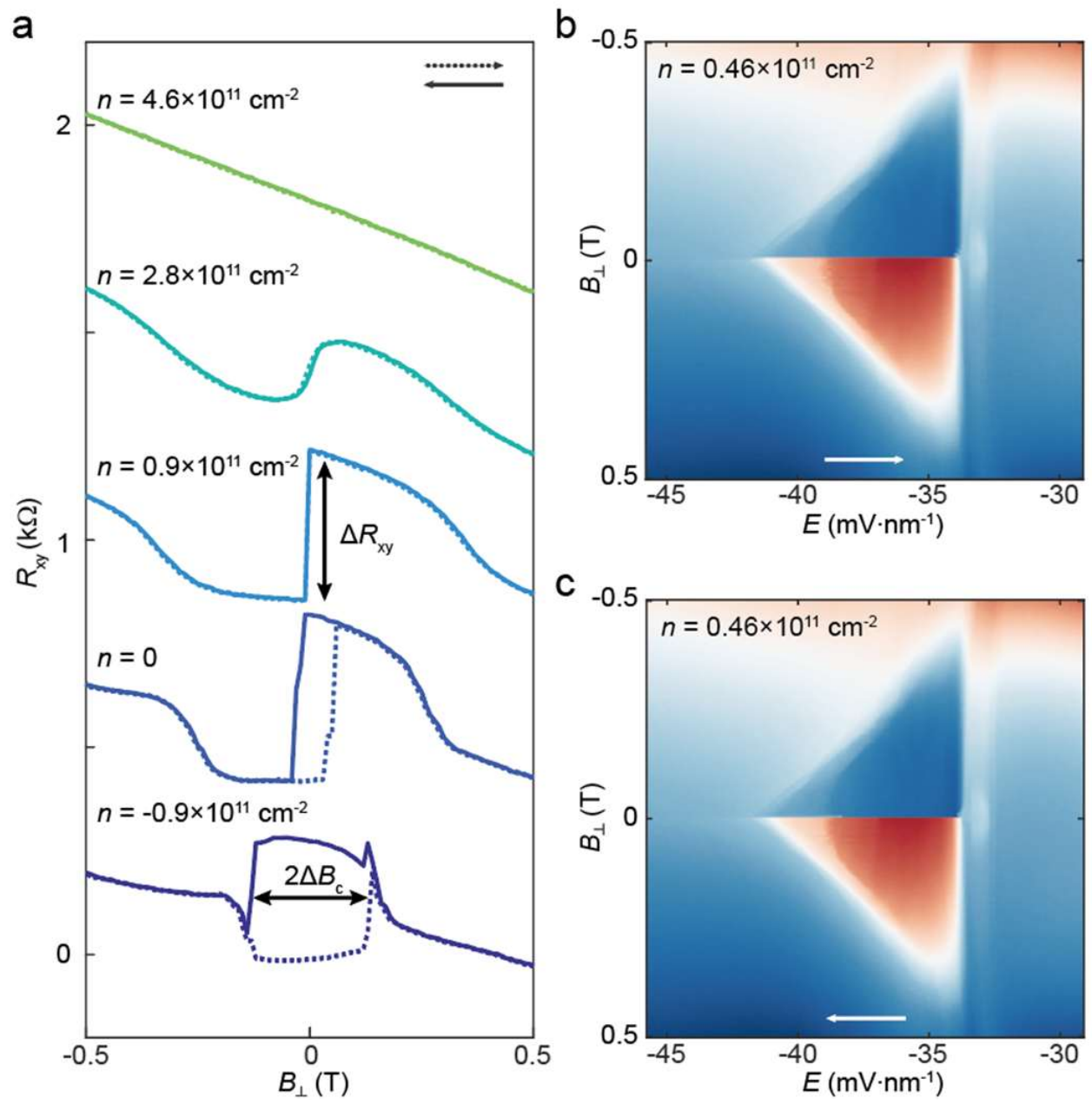

**Figure S13 | Carrier density driven ferromagnetic-paramagnetic transition. a,** Magnetic hysteresis of $R_{xy}$ recorded at different $n$ at $E$ = -36 mV·nm$^{-1}$ in the ferroelectric orbital magnetism regime. Dashed and solid curves represent the forward and backward scanning of $B_{\perp}$. Hall resistance difference $\Delta R_{xy}$ and coercive field $2\Delta B_c$ are defined by arrows. **b, c,** Color-coded $B_{\perp}$-dependent $E$ hysteresis scanning of the $R_{xy}$ at $n = 0.46 \times 10^{11}$·cm$^{-2}$, in the paramagnetic regime. White arrows in **e** to **f** represent the direction of scanning.

At $n$ = -0.9 × 10$^{11}$·cm$^{-2}$, the $R_{xy}$ signal shows a significant hysteresis Hall response with the sweeping of $B_{\perp}$, indicating the existence of ferromagnetism in this comet-like region with a coercive field $2\Delta B_c$ ~ 0.3 T and Hall resistance difference $\Delta R_{xy}$ = ~ 300 Ω. Increasing the carrier density to $n$ = 0 increases the $\Delta R_{xy}$ to ~ 400 Ω. While the $\Delta R_{xy}$ persists up to $n$ = 2.8 × 10$^{11}$·cm$^{-2}$, the strongly suppressed coercive field suggests a magnetic phase transition from ferromagnetism to paramagnetism.

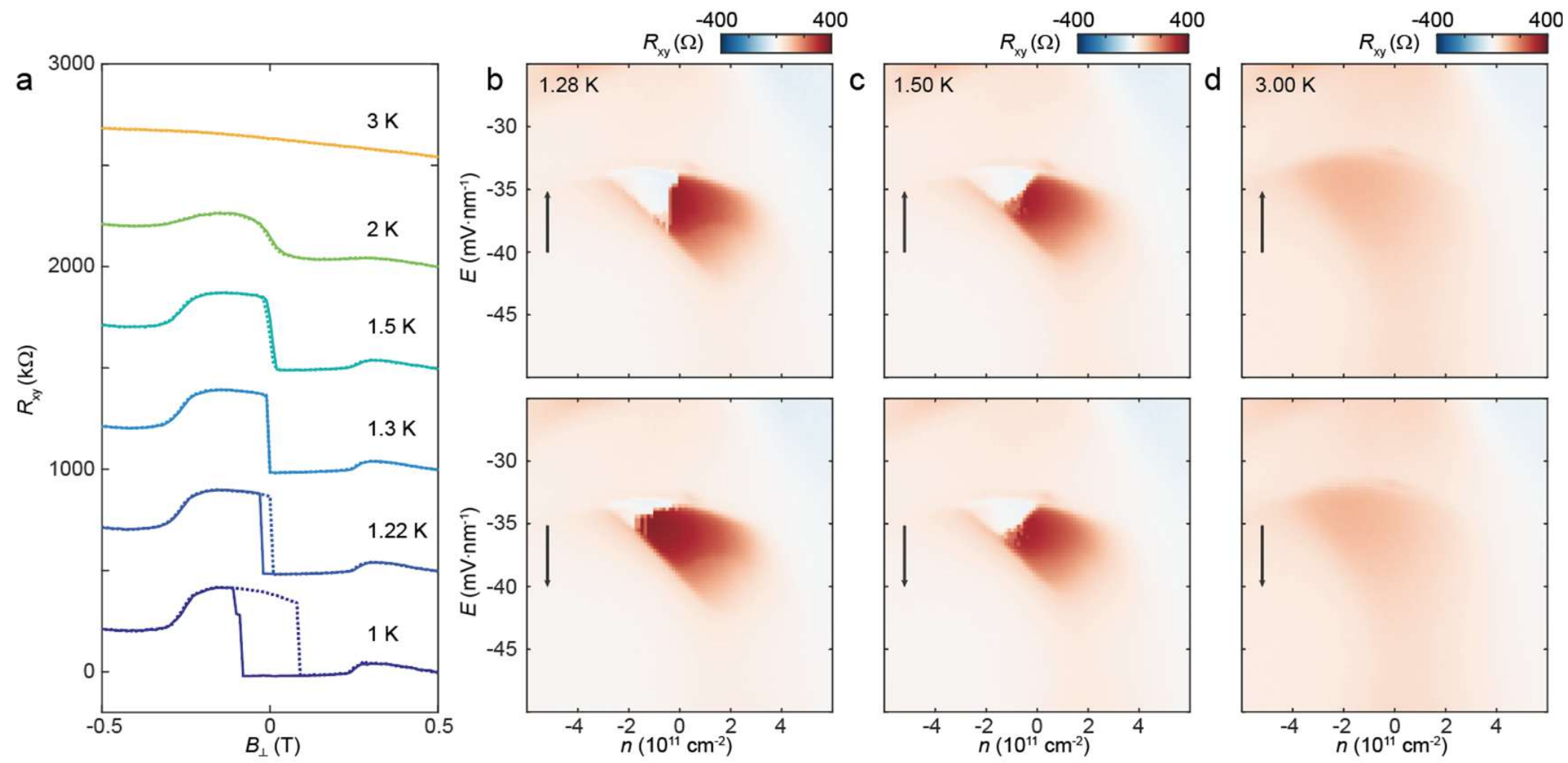

**Figure S14 | Temperature dependency of the ferroelectric orbital magnetism. a, b,** Temperature-dependent magnetic hysteresis of $R_{xy}$ at $n$ = -0.9 × $10^{11}$·cm$^{-2}$, $E$ = -36 mV·nm$^{-1}$. The result shows a vanished coercive field at 1.3 K, but sharp magnetization switches persist. At 2 K, the magnetization switch becomes soft, indicating a ferromagnetic to paramagnetic transition at this temperature. The signal of magnetization switches fully vanished at 3 K, suggesting that the orbital magnetism is fully suppressed at the elevated temperature. **b, c, d,** Color-coded temperature-dependent $E$ hysteresis of $R_{xy}$ measurements at 1.28, 1.5, and 3 K, respectively. Upper (lower) panels were acquired by sweeping $E$ backward (forward) along the arrow direction in each figure. Hysterisis vanishes at 1.5 K while the anomalous Hall signal persist up to 3 K, consistent with panel a.

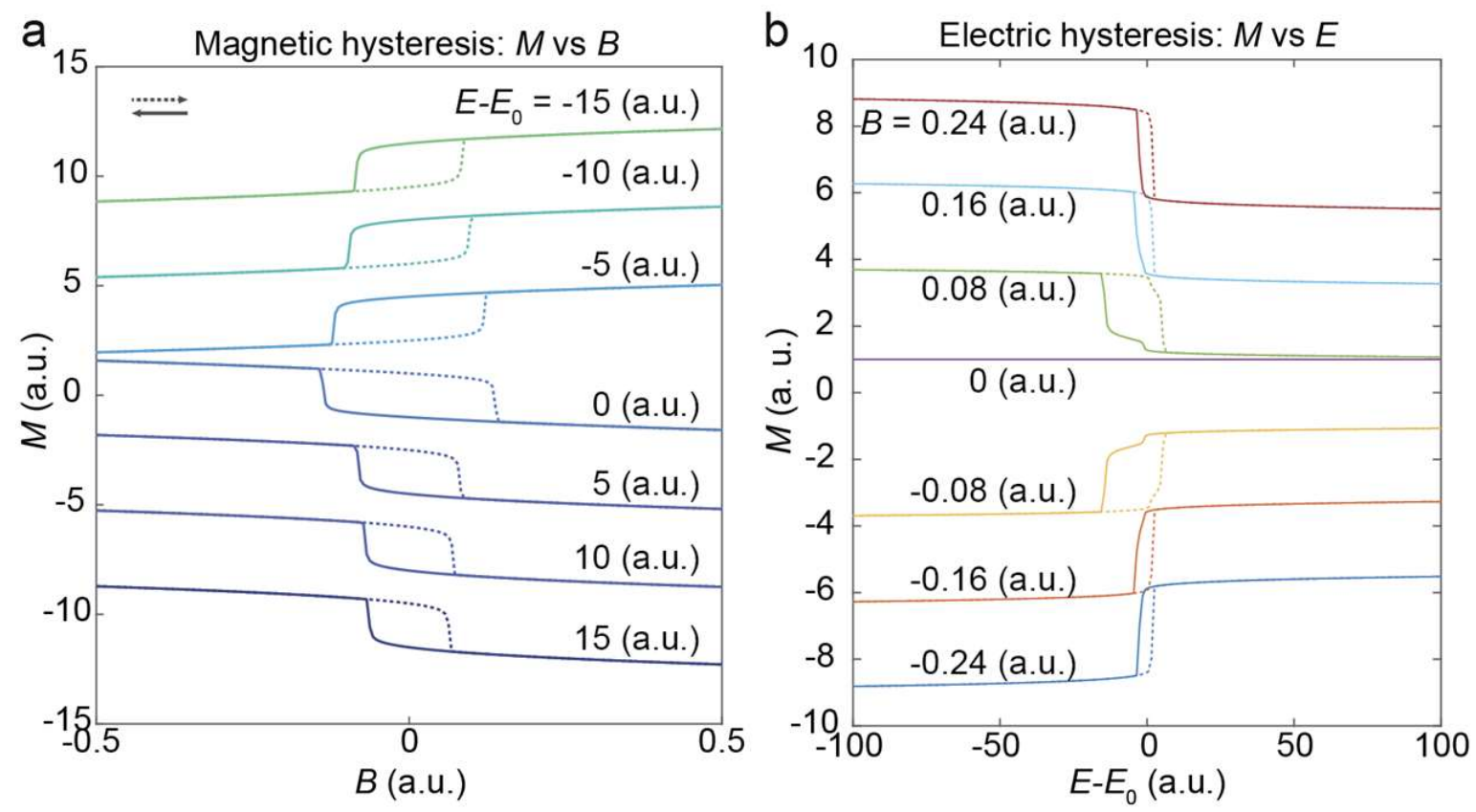

**Figure S15 | Simulation of the magnetic and electric hysteresis of ferro-electric orbital magnetism.** This simulation is performed numerically with the proposed free energy in Methods containing $F = -\lambda P \cdot M \cdot B$. Hysteresis is obtained by sweeping electric field or magnetic field while numerically relaxing P and M towards local free-energy minimum.

As shown in **a**, the polarity of the magnetic hysteresis reverses as the electric field is tuned through $E_0$. Furthermore, the hysteresis loop first enlarges, reaches a maximum near $E_0$, and then shrinks at larger fields, in agreement with experiment. In **b**, simulations of electric field sweeps similarly capture the main features observed in Figs. 4c, d. This free energy term may be interpreted as a layer-dependent orbital Zeeman coupling, though a microscopic understanding is beyond the scope of this phenomenological model and is left to future experimental and theoretical studies.

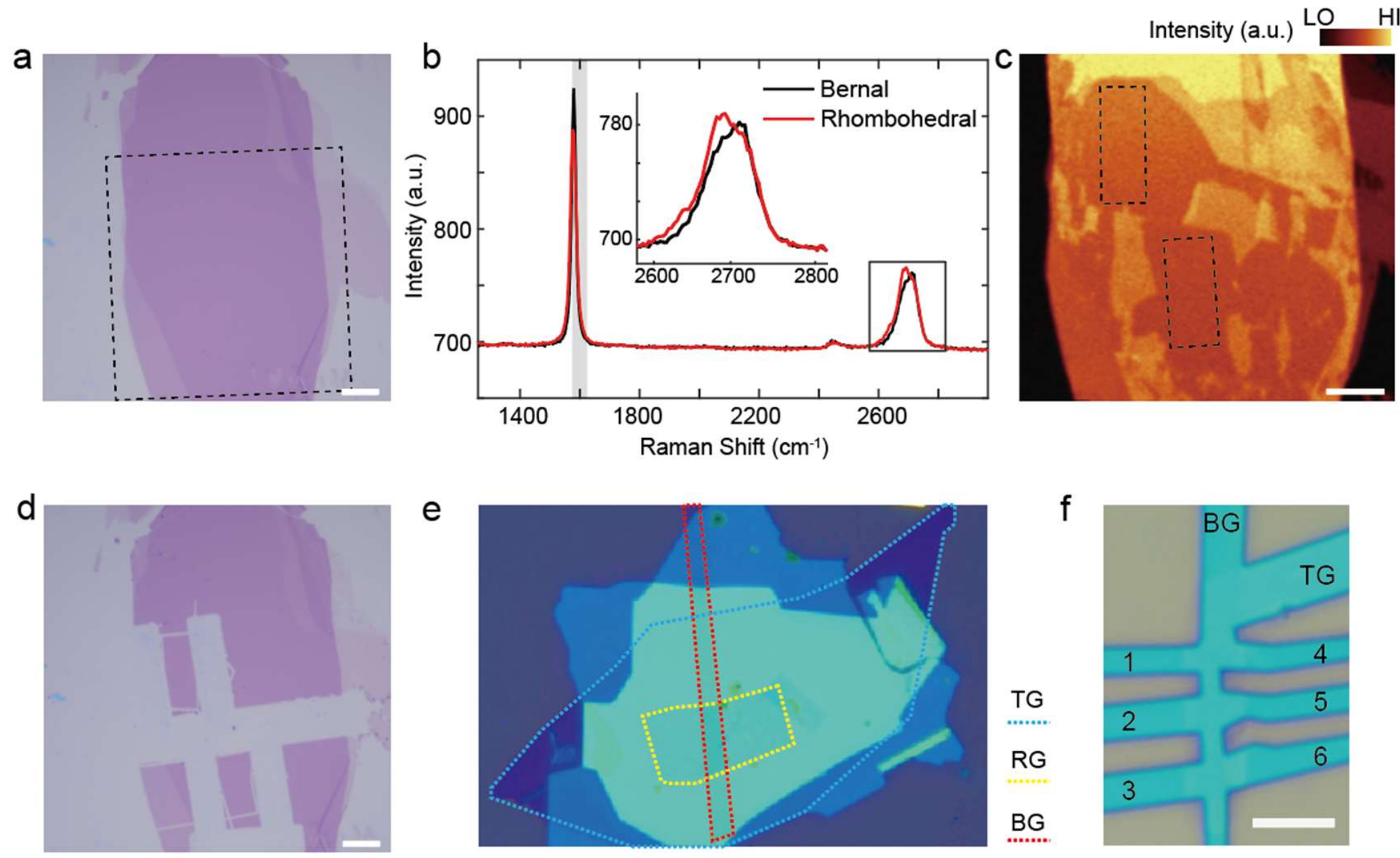

**Figure S16 | Determination, isolation, and stacking of rhombohedral graphene. a**, Optical microscope image of hexalayer graphene used for device fabrication. The thickness of the graphene layer is recognized by extracting the optical contrast between the $SiO_2$/Si substrate and the graphene layers. **b,** Typical large range Raman spectra of rhombohedral and Bernal hexalayer graphene, including both 2D band (~2700 $cm^{-1}$) and *G* band (~1600 $cm^{-1}$)[10]. While we still use the shape of a 2D band for the final determination of rhombohedral stacking order, we find that integrating half of the *G* band and mapping it out in real space provides a higher contrast between different domains. **c,** The Raman mapping extracted from integrating the G band (gray shadow in **b**). Black dashed boxes show the rhombohedral area that we isolated later by the AFM lithography method. **d,** The same flake as we have shown in **a** after AFM lithography. **e,** Optical microscope image of the final stack, where the top graphite gate (TG), rhombohedral graphene (RG), and bottom gate (BG) are highlighted by blue, yellow, and red dashed lines, respectively. **f,** Optical image after Hall bar device fabrication. Pins for electric transport measurements were marked by numbers 1 to 6.